\documentclass[traditabstract]{aa} 
\usepackage{graphicx}
\usepackage{txfonts}
\begin{document}
   \title{The effect of environment on star forming galaxies at redshift 1 -  \newline First insight from PACS}

\author{P. Popesso\inst{1}
\and
G. Rodighiero\inst{2}
\and
A. Saintonge\inst{1}
\and
P. Santini\inst{3}
\and
A. Grazian\inst{3}
\and
D. Lutz\inst{1}
\and
M. Brusa\inst{1}
\and 
B. Altieri\inst{4}
\and
P. Andreani\inst{5}
\and
H. Aussel\inst{6}
\and
S. Berta\inst{1}
\and
A. Bongiovanni\inst{7}
\and
A. Cava\inst{7}
\and
J. Cepa\inst{7}
\and
A. Cimatti\inst{8}
\and
E. Daddi\inst{6}
\and
H. Dominguez\inst{8}
\and
D. Elbaz\inst{6}
\and
N. F{\"o}rster Schreiber\inst{1}
\and
R. Genzel\inst{1}
\and
C. Gruppioni\inst{9}
\and
G. Magdis\inst{6}
\and
R. Maiolino\inst{3}
\and 
B. Magnelli\inst{1}
\and
R. Nordon\inst{1}
\and
A.M. P{\'e}rez Garc{\'i}a\inst{7}
\and 
A. Poglitsch\inst{1}
\and
F. Pozzi\inst{8}
\and
L. Riguccini\inst{6}
\and
M. Sanchez-Portal\inst{4}
\and
L. Shao\inst{1}
\and
E. Sturm\inst{1}
\and
L. Tacconi\inst{1}
\and 
I. Valtchanov\inst{4}
\and
E. Wieprecht\inst{1}
\and
M. Wetzstein\inst{1} \fnmsep\thanks{ Herschel is an ESA space observatory with science instruments provided by 
European-led Principal Investigator consortia and with important 
participation from NASA.}
}

   \institute{Max-Planck-Institut f\"{u}r Extraterrestrische Physik (MPE),
Postfach 1312, 85741 Garching, Germany.
\and
Dipartimento di Astronomia, Universit{\`a} di Padova, Vicolo dell'Osservatorio 3, 35122 Padova, Italy.
\and
INAF - Osservatorio Astronomico di Roma, via di Frascati 33, 00040 Monte Porzio Catone, Italy.
\and
Herschel Science Centre.
\and
ESO, Karl-Schwarzschild-Str. 2, D-85748 Garching, Germany.
\and
Laboratoire AIM, CEA/DSM-CNRS-Universit{\'e} Paris Diderot, IRFU/Service d'Astrophysique, 
B\^at.709, CEA-Saclay, 91191 Gif-sur-Yvette Cedex, France.
\and
Instituto de Astrof{\'i}sica de Canarias, 38205 La Laguna, Spain. 
\and
Dipartimento di Astronomia, Universit{\`a} di Bologna, Via Ranzani 1,
40127 Bologna, Italy.
\and
INAF-Osservatorio Astronomico di Bologna, via Ranzani 1, I-40127 Bologna, Italy.
}

   \date{Received September 1, 2010; accepted April 6, 2011}

 
  \abstract
  {We use deep 70, 100 and 160 $\mu$m observations taken with PACS, the Photodetector Array Camera and Spectrometer on board of Herschel, as part of the PACS Evolutionary Probe (PEP) guaranteed time, to study the relation between star formation rate and environment at redshift $\sim$ 1 in the GOODS-S and GOODS-N fields. We use the SDSS spectroscopic catalog to build the local analog and study the evolution of the star formation activity dependence on the environment. At $z \sim1$ we observe a reversal of the relation between star formation rate and local density, confirming the results based on Spitzer 24 $\mu$m data. However, due to the high accuracy provided by PACS in measuring the star formation rate also for AGN hosts, we identify in this class of objects the cause for the reversal of the density-SFR relation. Indeed,  AGN hosts favor high stellar masses, dense regions and high star formation rates. Without the AGN contribution the relation flattens consistently with respect to the local analog in the same range of star formation rates. As in the local universe, the specific star formation rate anti-correlates with the density. This is due to mass segregation both at high and low redshift. The contribution of AGN hosts does not affect this anti-correlation, since AGN hosts exhibit the same specific star formation rate as star forming galaxies at the same mass. The same global trends and AGN contribution is observed once the relations are studied per morphological type. We study the specific star formation rate vs stellar mass relation in three density regimes. Our data provides an indication that  at $M/M_{\odot} > 10^{11}$ the mean specific star formation rate tends to be higher at higher density, while the opposite trend is observed in the local SDSS star forming sample. }

   \maketitle
%

\section{Introduction}

One of the most fundamental correlations between the properties of galaxies in the local Universe is the so-called morphology (color) -density relation. This relation, quantified for the first time by Dressler (1980), shows that blue, star-forming, disk-dominated galaxies reside in lower density regions of the Universe than red, inactive elliptical galaxies.  The physical origin of the morphology density relation is still a subject of debate.  Much of the argument centers on whether the relation arises early on during the formation of the object, or whether it is caused by density-driven evolution.  Given the tight link between galaxy morphology, colors and instantaneous star formation rate (SFR), an alternative way to tackle this issue is to analyze the density-SFR relation. To shed light on how the environment affects the galaxy properties, these relations have to be studied at high redshift, when the galaxy formation processes are still undergoing.

The existence and the behavior of the density-SFR relation at redshift
$\sim 1$ is still matter of debate.  Kovac et al. (2009) show that
galaxy star-formation and color transformation rates are higher in the
groups than in lower density regions at $z \sim 1$. Elbaz et al (2007)
and Cooper et al. (2008) observe the reversal of the density-SFR
relation at $z\sim1$ in the GOODS and the DEEP2 fields, respectively,
using a spectroscopically defined density parameter. Caputi et al.
(2009) analyze the close environment, on 1 Mpc scales, of luminous
infrared galaxies (LIRGs, $L_{IR}=10^{11}-10^{12}$ $L_{\odot}$) and
ultra-luminous infrared galaxies (ULIRGs, $L_{IR}>10^{12}$
$L_{\odot}$) in the zCOSMOS dataset (Lilly et al. 2007), finding that
LIRGs at $0.6 < z < 1$ are more often found in overdense environment,
while ULIRGs prefer underdense regions. On the other hand, Feruglio et
al. (2010), using photometric redshifts to define the local galaxy
density, find no dependence of the SFR and LIRG fraction on
environment. The scenario is made even more complicated by the
interplay of mass and density. Indeed, Scodeggio et al. (2009) reveals that already at $z \sim 1$ mass and galaxy density are coupled with the most massive galaxies segregated in the most dense environment. Therefore, the
evidence for a clear density-SFR trend could be due to the different
contribution of massive and less massive galaxies favoring different
density regimes.  A first attempt to disentangle the mass-driven and
the environment driven evolution is provided by Peng et al. 2010,
based on the zCOSMOS and SDSS data. They argue that two distinct
processes are operating to affect the galaxy star formation activity,
"mass-quenching" and "environment-quenching".

A different approach to this issue is to study the SFR of galaxies
directly in high dense regions like clusters and groups. A rather long
list of cluster-related environmental processes can affect the SFRs of
galaxies. Some processes mainly affect the gaseous content of a
galaxy, such as the ram-pressure stripping (Gunn \& Gott 1972; Kenney
et al. 2004; van Gorkom 2004), re-accretion of the stripped gas
(Vollmer et al. 2001), turbulence and viscosity (e.g. Quilis et al.
2001), and starvation/strangulation (Larson et al. 1980).
Gravitational processes, which affect both the gaseous and the stellar
properties of a galaxy, range from low-velocity tidal interactions and
mergers (e.g. Mamon 1996; Barnes \& Hernquist 1996; Conselice 2006),
to high-velocity interactions between galaxies and/or clusters (Moore
et al. 1998, 1999; Struck 1999; Mihos 2004). Several of these
processes can affect the galaxy properties both by quenching or
enhancing the star formation activity. At low redshift it is well
recognized that the star formation is suppressed in groups and
clusters. The surrounding regions of groups and clusters have been
identified as the most likely environment where the star formation in
accreted galaxies is quenched (e.g. Abraham et al. 1996; Balogh et al.
1999; Pimbblet et al. 2002). Indeed, passive spirals (i.e. spiral
morphology but no star formation), which are considered objects in a
transition phase to an early type morphology, tend to be found in this
kind of intermediate density regions (e.g. Goto et al. 2003). The
situation seems to change towards higher redshift. At intermediate
redshift, Kodama et al. (2001) performed a wide-field imaging of the
CL0939 cluster at z= 0.41 and discovered that the color distribution
changes dramatically at the intermediate-density environment which
corresponds to groups/filaments. A very similar result was reported by
Tanaka et al. (2005) for the surrounding regions of higher redshift
clusters, CL0016 at z= 0.55.  Studies of clusters at redshift $z\sim
0.8$ (Koyama et al. 2008) show that star-forming activity is enhanced
in the intermediate-density cluster infalling region between
low-density general field and the high-density cluster core. More
recent studies of forming clusters (Tran et al. 2010, Hilton et al.
2010) show that towards higher redshift, $z\sim 1.4-1.6$, enhanced
star formation rates are observed also in the cluster core.  However,
these studies do not compare cluster and field galaxy star formation
activity at the same redshift. Thus, it is not clear whether this
enhancement of the cluster galaxy star formation activity is part of a
global enhancement at the peak of the SFR density at $z-1.5-2$ (Madau
et al. 1996) or is due to a real reversal of the density-SFR relation.

In this work we use PACS data to study the relation between the
galaxy SFR and the environment at redshift $\sim 1$. In order to
disentangle the mass segregation from the environmental effects, we
dissect in mass bins the instantaneous galaxy density-SFR relation and
in density bins the specific SFR-stellar mass relation.  To achieve
this goal we need deep IR (dominating the bolometric luminosity at
high SFR) data to get a reliable estimate of the SFR. We also require
a highly complete spectroscopic coverage to estimate the local galaxy
density.  Thanks to the observations carried with PACS (Photodetector
Array and Camera Spectrometer, Poglitsch et al. 2010, Pilbrat et al.
2010) as part of the PACS Evolutionary Probe (PEP) GT Program, we are
able to meet these requirements. Indeed, PACS with PEP provides at the
moment the deepest far-infrared survey at 70, 100 and 160 $\mu$m of
the GOODS fields, for which an extremely rich multiwavelength dataset
and extremely highly complete spectroscopic catalog are available.
Moreover, the GOODS fields show the presence of several intermediate
density regions, such as groups and filaments, at $z\sim 1$, where the
enhancement of the star formation activity should be found.  We use
also the SDSS DR7 spectroscopic catalog to relate the $H\alpha$ based
SFR to the local galaxy density at $z\sim 0$ in order to evaluate the
evolution of the density-SFR relation. The paper is organized as
follows. In Section 1 we describe the data and the sample selection.
In Section 2 we describe the methods applied to estimate the galaxy
properties. In Section 3 we provide our results about the relation of
the SFR and the local environment at redshift $\sim 1$. In Section 4
we provide the evolutionary analysis of the density-SFR relation. In
Section 5 we list our conclusions.

We adopt $\Omega_{\Lambda}=0.73$, $\Omega_{m}=0.27$ and $H_0=75 \rm{km}{s}^{-1}\rm{Mpc}^{-1}$ throughout this paper.


\section{Data and sample selection}

\begin{figure}
   \centering
  \includegraphics[width=8cm]{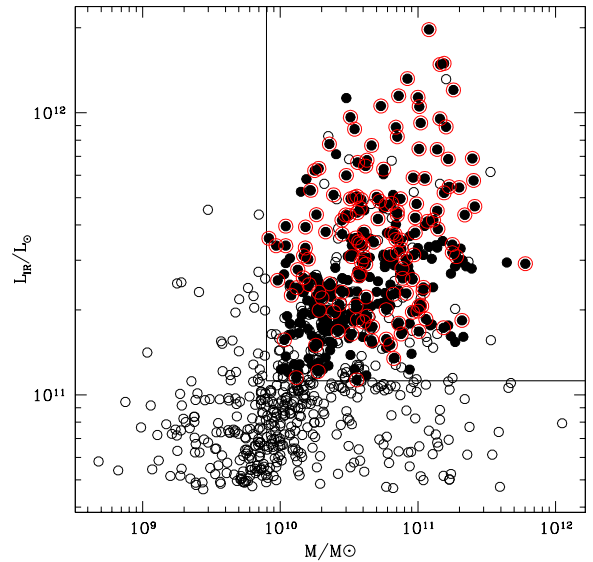}
\caption{Bolometric IR luminosity - stellar mass relation. Empty circles in the figure are MIPS 24 $\mu$m detections without spectroscopic information. Filled circles in the figure are MIPS 24 $\mu$m detections with spectroscopic information. Red circles identify MIPS sources with PACS detection. The horizontal and vertical lines shows the $M/M_{\odot}=8\times 10^9$ and $L_{IR}/L_{\odot}=10^{11}$ cuts corresponding to our selection criteria at $0.7 < z < 1.1$. }
         \label{figura1}
   \end{figure}

   \begin{figure*}
   \centering
\includegraphics[width=6cm]{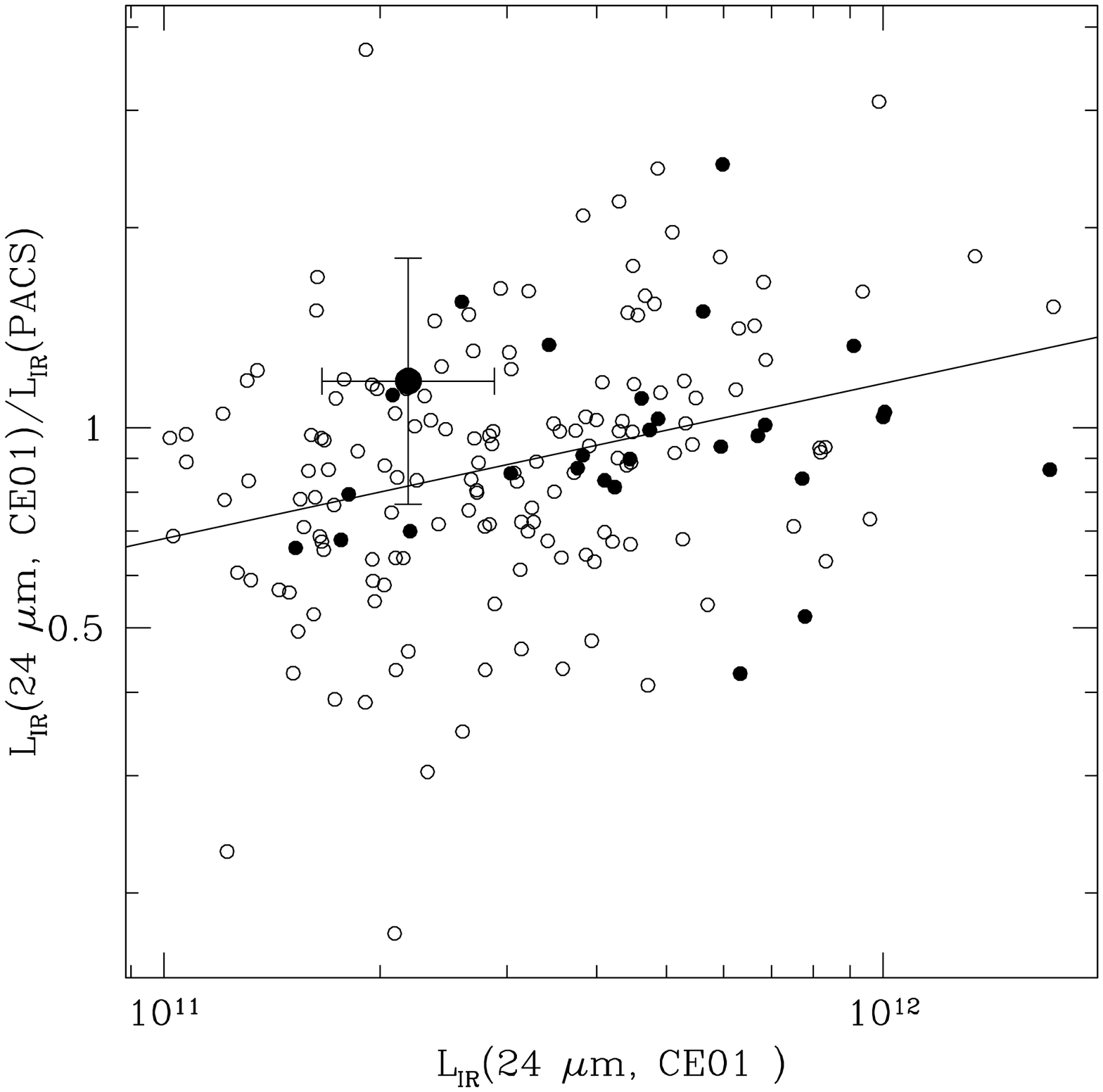}
\includegraphics[width=6cm]{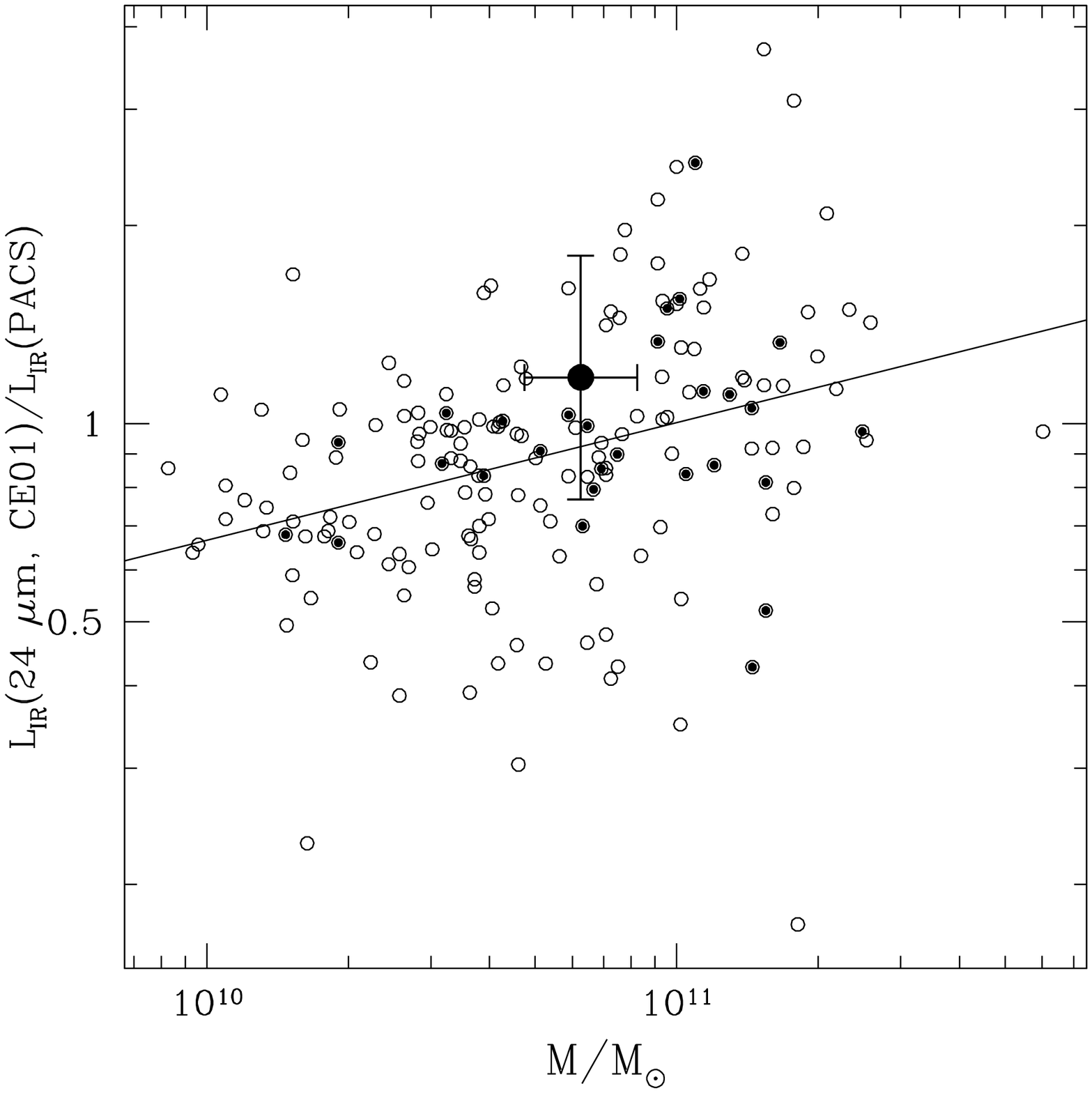}
\includegraphics[width=6cm]{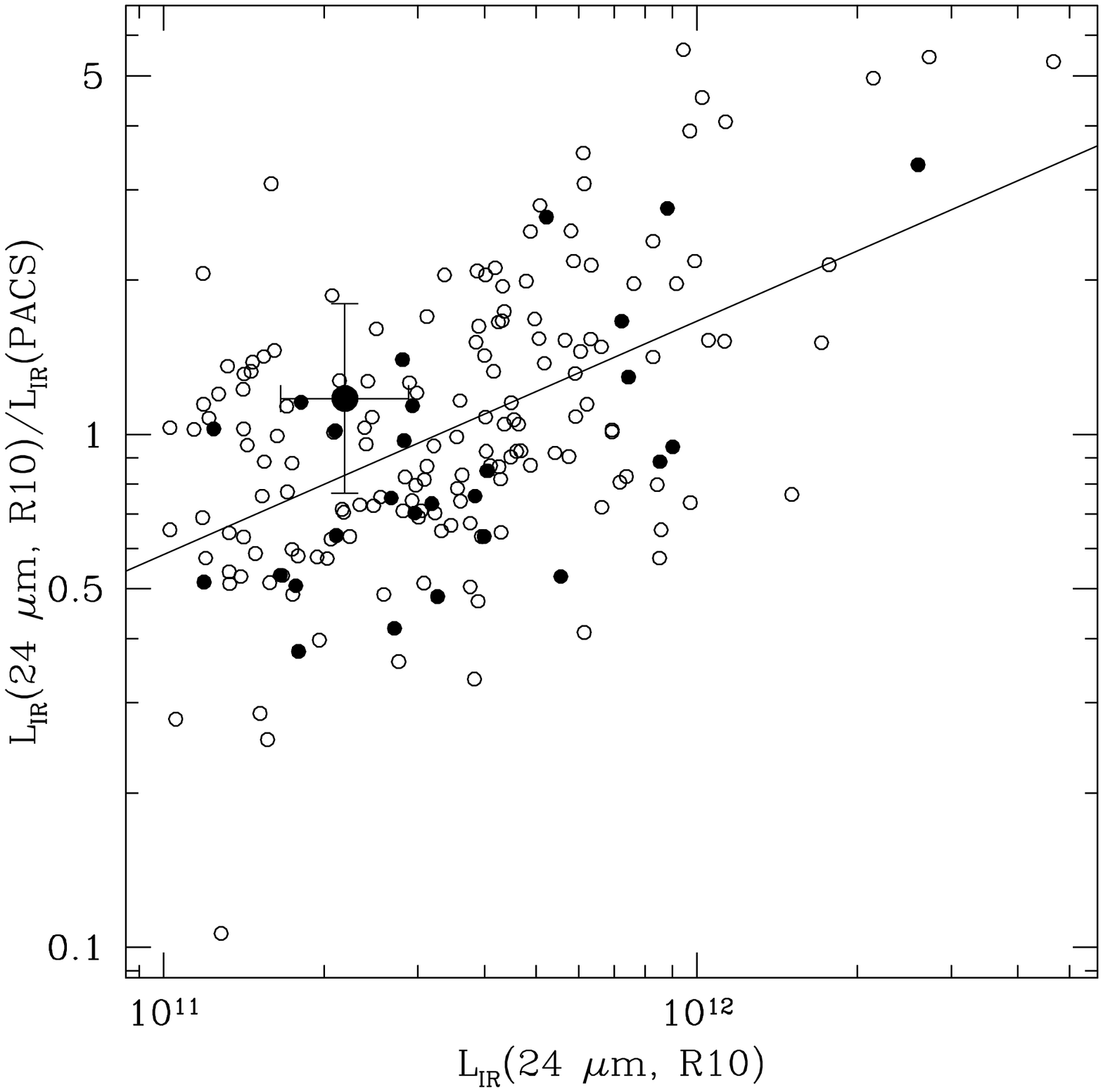}
\caption{$L_{IR,24 {\mu}m}/L_{IR, PACS}$ ratio versus  $L_{IR,24 {\mu}m}$ obtained using the CE01 method, in the left panel. $L_{IR,24 {\mu}m}/L_{IR, PACS}$ ratio versus the $M/M_{\odot}$ using the CE01 method, in the central panel. $L_{IR,24 {\mu}m}/L_{IR, PACS}$ ratio versus the $L_{IR,24 {\mu}m}$ obtained using the R10 method, in the right panel. In all panels empty circles are normal star forming galaxies in our sample with both a MIPS 24 $\mu$m and PACS 100 or 160 $\mu$m detection. Filled points are low luminosity obscured AGNs  with both a MIPS 24 $\mu$m and PACS 100 or 160 $\mu$m detection. The point with error bar is the result of the stacking analysis in the GOODS-N 160 $\mu$m map of the  obscured AGNs of our sample with only MIPS detection. }
         \label{figura2}
   \end{figure*}

\subsection{PACS data}

In this work we use the PACS data of the PACS Evolutionary Probe (PEP)
guaranteed time observations collected in the GOODS-N and GOODS-S
fields. The GOODS-N 30h observations were carried out in medium speed
(20''/s) scan map mode at 100 and 160 $\mu$m during the science
demonstration phase of the Herschel mission. The GOODS-S 113 h
observations were taken in medium speed scan map mode at 70, 100 and
160 $\mu$m in December/January (2009-2010). Data have been processed through the
standard PACS data reduction pipeline in the Hipe environment (Ott et
al. 2006), with the addition of some custom procedures aimed at
removal of instrument artifacts.

Glitch removal is based on the Multiresolution Median Transform, a
method developed by Starck \& Murthag (1998) for the detection of
faint sources in ISOCAM data. The method relies on the evidence that
signal due to a real source and a glitch, respectively, shows
different signatures in the pixel time-line. These features are
identified using a multiscale transform separating the various
frequencies of the signal. Once the glitch components are identified,
they are replaced by interpolated value in the pixel time-line.

PACS photometer observation exhibit 1/f noise (see Lutz et al. in prep. for a more detailed explanation). In oder to remove the bulk of the noise we apply a ``running-box'' median filter to the pixel time-line. The source positions are masked in order to avoid source flux subtraction. The source mask is created by masking all pixels with S/N above a given threshold in a preliminary map.  The source mask is optimized by iterating the procedure.

The blue channel (70 and 100 $\mu$m) data show clear interference patterns which
 are not ascribable to the 1/f noise. In order to remove those features, the individual scan legs of each observation have been visually checked and flagged when needed. 

 Before the map reconstruction, offsets and errors in the pointing
 accuracy of the Herschel satellite have been corrected by
 re-centering the data on a grid of known 24 $\mu$m sources. Scan legs
 of different scan directions and repetition numbers have been
 projected in separate maps. The 24 $\mu$m sources were then stacked
 in the individual maps to compute the average $\rm{\Delta x}$ and
 $\rm{\Delta y}$ offsets to be applied to the given set of scan legs.

 The map reconstruction was done via a drizzle method (Fruchter \&
 Hook, 2002). Given the high data redundancy in the GOODS fields, we
 chose a very small drop size ($pixfrac=1/8$ with a final output pixel
 scale of 2'' at 70 and 100 $\mu$m and 3'' at 160 $\mu$m) to reduce
 the correlated noise in the final map (see Fruchter \& Hook, 2002 for
 a detailed discussion). The individual maps were then coadded,
 weighting pixel contributions by the effective exposure. The final
 error map is derived as the standard deviation of the weighted mean
 in each map pixel.  Due to the map projection and residual 1/f noise,
 the output pixels show partial correlation. A correlation correction
 factor is calculated in different subareas of the final map under the
 assumption that it depends only on relative pixel positions. The
 correlation turned out to be rather uniform across the final map.
 Thus, a mean correlated noise correction factor is used in our noise
 estimation.

 The Point Spread Function (PSF) were extracted from the final science
 maps, and have a FWHM of $\sim$ 7.1, $\sim$ 7.5 and $\sim$ 11 arcsec
 in the 70, 100 $\mu$m and 160 $\mu$m bands, respectively. The aperture
 correction is based on the calibration observation of the Vesta
 asteroid. Particular care was taken in calibrating the derived flux
 densities to account for the known flux overestimation in the used
 Hipe version by factors 1.02, 1.09 and 1.29 at 70, 100 $\mu$m and 160
 $\mu$m, respectively as described in the PACS scan map release note
 PICC-ME-TN-035. As outlined in the note, the absolute flux accuracy
 is within 10 \% at 70 and 100 $\mu$m and better than 20\% at 160 $\mu$m.

 PACS catalogs were created using two different approaches: a blind
 extraction using the Starfinder PSF-fitting code (Diolaiti et al.
 2000) and a guided extraction using 24 $\mu$m priors. Flux
 reliability, incompleteness and spurious source fraction were
 estimated and tested via Monte Carlo simulations.  We created 500
 images of the individual fields by adding $\sim$ 20 artificial
 sources each for a total of 10000 sources. Input and output fluxes
 are consistent within a few percent. The completeness is defined as
 the fraction of sources detected with photometric accuracy of at
 least 50\% (Papovich et al. 2004). Spurious sources are defined as
 those detected at 3$\sigma$ flux level and with an input flux lower
 than 3$\sigma$. The GOODS-N observations reach the 3$\sigma$ level at
 $\sim$3.0 mJy and $\sim$5.7 mJy at 100 and 160 $\mu$m, respectively.
 The observations in GOODS-S reach the 3$\sigma$ limit at
 $\sim$ 1.2, 1.2, 2.4 mJy at 70, 100, 160 $\mu$m, respectively. In order to test our
 results against biases, we performed the analysis with both catalogs
 to compare the results. The use of different catalogs does not lead
 to any significant difference. Thus, hereafter we show only the
 results based on the 24 $\mu$m prior catalog. The final catalog
 comprises all PACS detections down to the 3$\sigma$ level. We reach a
 completeness level of $\sim$ 70\% at the 5$\sigma$ level and of 35\%
 at the 3$\sigma$ level. The spurious contamination is about 10\%.

\subsection{Auxiliary data}

The PACS catalogs are matched to highly reliable multiwavelength
catalogs in both fields. For GOODS-N the PEP Team created a
PSF-matched database adopting the Grazian et al. (2006, ConvPhot)
approach, including UV (GALEX), optical (HST), near-infrared
(FLAMINGOS) and infrared (Spitzer IRAC and MIPS) data, which includes
also the spectroscopic redshifts collected by Barger et al. (2008). We
include in the catalog also the X-ray rest-frame luminosity of
Alexander et al (2003) based on the Chandra 2 Ms observations of
GOODS-N. For GOODS-S we use the second version of the PSF-matched
GOODS-MUSIC catalog (Santini et al. 2009), which provides U band (2.2
ESO and VLT-VIMOS), optical (HST), near-infrared (VLT-ISAAC) and
infrared (Spitzer IRAC and MIPS) data. We matched the GOODS-MUSIC
catalog to the GOODS-S spectroscopic redshift master catalog of
Balestra et al. (2010) after removal of multiple entries for the same
source.  The GOODS-MUSIC catalog is matched also to the Luo et al.
(2008) X-ray catalog based on the 2 Ms observation of GOODS-S.

\subsection{The sample selection}

For the purpose of the paper we need to define two different catalogs:
the star forming galaxy sample and the galaxy sample to use for building the
density field around each star forming galaxy. In this section we
describe how we select the star forming galaxy sample at high and low
redshift.

\subsubsection{Star forming galaxies at z$\sim$ 1}
The PACS GOODS-N and GOODS-S observations are quite different in terms
of depth. The GOODS-S maps are twice as deep as the GOODS-N data. In
order to harmonize the two datasets without loosing in depth, we
complement the PACS data with MIPS 24 $\mu$m data, which reach the
3$\sigma$ level at $\sim$ 25 $\mu$Jy in both fields (Elbaz et al.
2007).  This solves also any incompleteness issue of the GOODS-N and
GOODS-S PACS samples between the 3 and 5$\sigma$ flux levels. Given
the need for spectroscopic redshifts to reliably estimate the local
galaxy density, we use a subsample of the combined PACS+MIPS sample
with high quality spectroscopic redshift. The analysis of the
spectroscopic completeness shows that this is above $\sim$ 60 \% in
any flux bin down to 80 $\mu$Jy for both GOODS-N and GOODS-S PACS
sample. Below 80 $\mu$Jy the spectroscopic completeness per flux bin in
the 24 $\mu$m sample drops down quickly to $\sim$ 35 \%. Thus, we
apply a 80 $\mu$Jy cut in the 24 $\mu$m sample. To ensure mass
completeness we apply an additional cut at [4.5 $\mu$m] $< $ 23 mag
(Mancini et al. 2009).

We chose the redshift range $0.7 < z < 1.1$ in order to sample the two
main large scale structures in the fields: a low mass cluster ($ M
\sim 10^{14} M_{\odot}$ ) at $z=0.736$ (Popesso et al. 2009) in
GOODS-S and a big group at $z=1.016$ (Elbaz et al. 2007) in GOODS-N.
Fig. \ref{figura1} shows the PACS+MIPS sources identified by our
selection criteria in the $M/M_{\odot}-L_{IR}$ plane. The horizontal
and vertical lines shows the $M/M_{\odot}=8\times 10^9$ and
$L_{IR}/L_{\odot}=10^{11}$ cuts corresponding to our selection
criteria at $0.7 < z < 1.1$.The mass limit is estimated as
    described in Rodighiero et al. (2010): we computed the
    mass-completeness thresholds as a function of redshift for our
    IRAC 4.5 $\mu$m-selected sample (mag 4.5 $\mu$m $<$ 23.0, AB),
    derived from synthetic stellar population models as described in
    Mancini et al. (2009). We used a variety of galaxy models from the
    templates of Bruzual \& Charlot (2003), with a Salpeter IMF, and
    different ages, and dust extinction parameters (E(B-V)). We
    decided to adopt the most conservative mass-completeness limit,
    above which even the oldest (2 Gyr) and highly extincted (E(B-V) $=$
    0.8) star-forming galaxy population would be entirely recovered.
The filled points in the figure refer to galaxies with spectroscopic
redshift. The spectroscopic completeness is fairly uniform in both
mass and $L_{IR}$ bins. Red circles identify PACS detections, which
dominate the $L_{IR}$ bright end. In addition, the cut at
$L_{IR}/L_{\odot}=10^{11}$ nicely separates the star forming and
passive galaxies at the high mass end.

We identify also 60 X-ray detected sources in our sample, 40 in
GOODS-N and 20 in GOODS-S. 49 of these sources are identified as AGN
following the criteria of Brusa et al. (2009). 3 out of 49 AGNs are
broad line AGNs. We exclude those sources from our sample since in
those cases the AGN contribution dominates the SED at optical and IRAC
wavelengths, thus, not allowing a reliable estimate of the host galaxy
stellar mass (see Merloni et al. 2010).  The dominant AGN contribution
at MIR wavelength prevents also a reliable estimate of the host galaxy
star formation rate.  The remaining 46 AGNs are classified as obscured
AGNs, the dominant population in the GOODS fields as outlined in
Alexander et al. (2003) and Luo et al. (2008).  These sources are all
at luminosities below that of typical QSOs ($L_X>10^{44}$ erg
s$^{-1}$).  Moderate-luminosity AGN ($L_X\sim10^{43}$ erg s$^{-1}$)
have been shown to be ideal laboratories for the study of their host
galaxies due to their low optical brightness which in many cases is
due to their nuclear obscuration (Tozzi et al. 2006, Mainieri et al.
2007).  In particular, as shown in Merloni et al. (2010), an obscured
AGN contributes in a negligible way to the host galaxy SED at shorter
wavelengths, thus allowing a reliable estimate of the host galaxy
stellar mass. 22 out of the 46 obscured AGNs in GOODS fields have only
a 24 $\mu$m detection. MIR wavelength can be affected by the AGN
contribution and can provide overestimated estimates of the
instantaneous star formation rate.  Instead, the rest-frame far
infrared seems to be dominated by the host (Netzer et al. 2007, Lutz
et al. 2010). In order to check whether the MIR flux provides a
reliable estimate of the host galaxy star formation rate, we compare
the $L_{IR}$ extrapolated from the 24 $\mu$m flux versus the $L_{IR}$
given by the PACS FIR flux, for the sources which have both
detections. Fig. \ref{figura2} shows the ratio $L_{IR,24
  {\mu}m} /L_{IR, PACS}$ versus the $L_{IR,24 {\mu}m}$. The
trend of the relation is discussed in the next session, but we point
out here that, with the exclusion of the broad lines AGNs, the
remaining sources lie on the same relation and show the same scatter
as the normal star forming galaxies. To check the reliability of the
24 $\mu$m extrapolation, we stacked in the GOODS-N 160 $\mu$m map the
17 AGNs without PACS detection. The mean flux is then converted into
$L_{IR}$ by using the empirical $L_{160 {\mu}m} - L_{IR}$ relation
found in our dataset. The stacked point gives a ratio $L_{IR,24{\mu}m} /L_{IR, PACS}$ of 1 and it is consistent within $1\sigma$ with the relation obtained with the individual detections. We conclude that in the redshift range of interest $L_{IR}$ extrapolated from the 24 $\mu$m flux can provide a reliable estimate of the host galaxy star formation rate.

After the exclusion of broad line AGNs, we are left  with 326 sources, of which 185  are PACS sources.

\subsubsection{The $z \sim 0$ counterpart: the Sloan star forming galaxy sample}
The local galaxy sample is drawn from the SDSS DR7 spectroscopic
database (Abazajian et al. 2009). The spectroscopic component of the
SDSS survey is carried out using two fiber-fed double spectrographs,
covering the wavelength range 3800-9200 $Å$, over 4098 pixels. They
have a resolution varying between 1850 and 2200, and together they are
fed by 640 fibers, each with an entrance diameter of 3 arcsec. The
fibers are manually plugged into plates inserted into the focal plane;
the mapping of fibers to plates is carried out by a tiling algorithm
(Blanton et al. 2003) that optimizes observing efficiency in the
presence of large-scale structure.

Using the selection criteria of Yasuda et al. (2001) we identify
813287 unique galaxies in the SDSS galaxy spectroscopic sample.  We
match this sample to the catalog provided by the MPA-JHU DR7 release
of spectrum measurements ($http://www.mpa-garching.mpg.de/SDSS/DR/$).
This catalog provides stellar mass estimates based on Kauffmann et al.
(2003) and Salim et al. (2007). The star formation rate and specific
star formation rate are based on Brinchman et al. (2004, hereafter
B04, we refer to this work for further detail of the SFR and sSFR
estimation). The current version of the catalog provides mainly
    $H\alpha$-derived SFR. An estimate of the SFR is provided also for
    galaxies without emission lines. This estimate is based on D4000
    index. Since no calibration is provided for this indicator, we
    limit our sample to the galaxies with emission lines and
    $H\alpha$-derived SFR. The total SFR and sSFR is obtained by
    correcting the fiber estimates. The extinction correction is based on the $H\alpha/H\beta$ ratio. The aperture corrections are done
    by fitting the photometry of the outer regions of the galaxies
    with models and should remove the systematic overestimate of total
    SFR in certain classes of galaxies identified by Salim et al
    (2007). B04 divide the galaxy sample into several classes on the
    basis of the BPT diagram (Baldwin et al. 1981). The following
    galaxy categories are identified: star-forming galaxies, low S/N star forming galaxies, composite
    galaxies, AGNs, and unclassifiable objects.  Since the AGN
    contribution to the $H\alpha$ flux can lead to a SFR
    overestimation, we exclude this class of objects from the sample. Moreover, we limit our sample to the galaxies classified as
    star forming and low S/N star forming. We exclude the unclassifiable objects, which, according to B04 are galaxies with extremely low S/N or absent $H\alpha$ emission and with very uncertain SFR estimate.

    We select a subsample of galaxies in the redshift slice $0.01 < z<
    0.1$  and with the same stellar mass cut used for the high redshift
    sample, $M/M_{\odot}=8\times 10^9$, for a total number of 97244
    star forming galaxies.

\begin{figure}
   \centering
  \includegraphics[width=8cm]{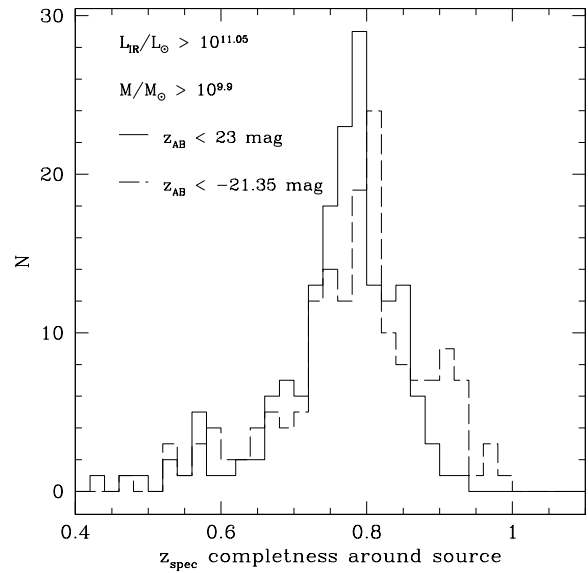}
\caption{Distribution of the completeness parameter calculated around each source. The solid line shows the histogram of the completeness parameter based on a galaxy sample selected with an apparent magnitude cut at $z_{AB} < 23$ mag. The dashed histogram shows the distribution of the completeness parameter based on a galaxy sample selected with an absolute magnitude cut at $z_{AB} < -21.35$ mag.}
         \label{figura3}
   \end{figure}

\begin{figure}
   \centering
 \includegraphics[width=7.5cm]{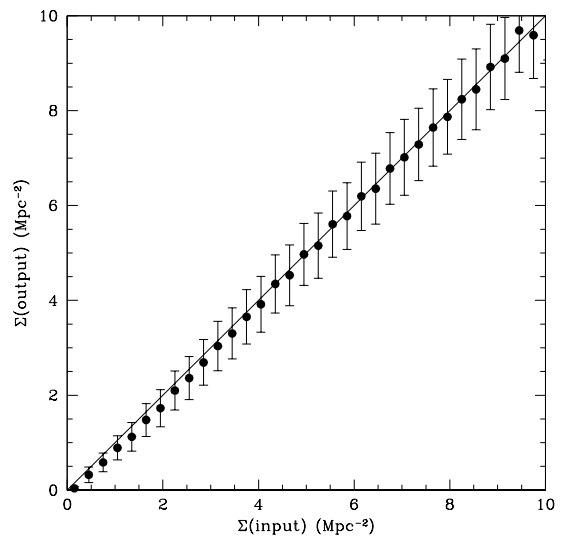}
\caption{Comparison of the density parameter calculated assuming 80\% incompleteness and corrected for that($\Sigma(output)$)  versus the actual value ($\Sigma(input)$) in a light-cone of the Millennium simulation (Springel et al. 2005). }
         \label{figura3a}
   \end{figure}

\section{Galaxy properties estimate}

\subsection{The local galaxy density parameter}

To reconstruct the density field around each star forming
    galaxy in the GOODS-N and GOODS-S fields we use all galaxies
    with spectroscopic redshift in the range $0.7 -1.1$ and with
    $z_{AB} < 23$ mag to ensure a rather high spectroscopic
    completeness level (Popesso et al. 2009).

    We compute the projected local galaxy density, $\Sigma$, by
    counting all galaxies located inside a radius of $0.75$ Mpc and
    within a fixed velocity interval of $\Delta \rm{v}=3000$
    $\rm{km}{s}^{-1}$, 3 times the typical velocity dispersion of
    massive clusters ($\sigma_v \sim 1000$ $\rm{km}{s}^{-1}$), around
    each galaxy. The key ingredient for building a reliable density
    field is a very high and spatially uniform spectroscopic coverage.
    For this purpose we measure the spectroscopic completeness around
    each source as the ratio between the number of objects with
    spectroscopic redshift and the number of all sources with and
    without redshift, at $z_{AB} < 23 $ mag in the cylinder along the
    line of sight of the considered source and with a radius
    corresponding to 0.75 Mpc at the redshift of the considered
    source. As shown in Fig.  \ref{figura3}, the distribution of the
    completeness parameter is very narrow with a rms of 0.04 around the
    median value of 77\%. This ensures that the spectroscopic coverage
    of the fields is very high and rather uniform. However, number
    densities measured within an apparent magnitude limit exhibit a
    strong redshift bias. Indeed, an apparent magnitude limit samples
    different regions of the rest frame luminosity function (LF) at
    different redshifts: the lower the redshift, the larger the region
    of the LF sampled, thus the larger the number of galaxies measured
    within the cylinder. For this reason, we prefer to measure the
    number densities at an absolute magnitude cut. We set the cut to
    the absolute magnitude corresponding to $z_{AB} = 23 $ at the
    highest redshift of our sample, $z=1.1$, which is $z_{AB} =
    -21.35$ .  The local galaxy density and the local completeness in
    this case are estimated as described above with the difference
    that galaxies are counted down to the
    $m(z)=-21.35+5log(dl)+5log(h)+k(z)+25$, where $z$ is the redshift
    of the reference star forming galaxy at the center, $dl$ is its distance
    luminosity and $k(z)$ is the k correction with respect to $z=1.1$.
    The k correction is estimated through the $Kcorrect$ code of
    Blanton et al.  (2007, version $\rm{v4\_1\_4}$). The distribution of the completeness
    parameter is as narrow as the one obtained with the apparent
    magnitude cut, with a somewhat higher peak at 80\% $\pm$ 0.05, as
    shown in Fig.  \ref{figura3} (dashed histogram).  Thus, our
    absolute magnitude cut ensures the highest completeness level up
    to the highest redshift of our sample without biasing the density
    parameter.  We use our local completeness measure to correct the
    local galaxy density to retrieve the actual density field. This is
    done under the assumption that the incompleteness in the cylinder
    along the line of sight around each galaxy is the same as in the
    portion of the same cylinder within $\Delta \rm{v}=3000$ $\rm{km}{s}^{-1}$ from the
    redshift of the galaxy.

    The reliability of the method is also tested on a sample of
    simulated galaxies in lightcones built from the Millenium
    simulation (Kitzbichler \& White 2007, Springel et al. 2005) in
    the same redshift range considered in this analysis and at $M_B <
    -20.75$ mag, corresponding to our absolute magnitude cut at $z \sim 1$. We assume a mean spectroscopic completeness of 80\%
    and extract randomly a corresponding percentage of galaxies in the
    lightcone. We apply our method to estimate the projected local
    galaxy density and compare with the actual value in Fig.
    \ref{figura3a}.  The corrected projected local galaxy density
    based on 80\% completeness and the real density lie on average on
    the 1 to 1 line. The errors in the figure are estimated through a
    jackknife technique. According to the mock catalog result, the
    average uncertainty in our density estimate is about 15\%, with a
    maximum of $\sim$ 23\% at the minimum density and a minimum of
    $\sim$ 7\% at the maximum density. This shows that using our
    approach we trace reliably the actual density field.

We compared this local density estimator with similar methods. Elbaz
et al. (2007) vary the redshift interval in order to sample the same
comoving value, 90 $\rm{Mpc}^3$, around each galaxy. Instead, as in
Poggianti et al. (2010) we use proper quantities. This is motivated by
the fact that, in order to study the dependence of galaxy properties
on the density of the local environment, what matters are
gravitational and vicinity effects, and therefore proper distances
between galaxies. However, given the narrow redshift slice considered
here, we see that our estimates of the surface and volume densities
are consistent within the errors with the quantities derived at a
fixed comoving volume. Our approach is also consistent with the
    method of Elbaz et al. (2007) once we use the same apparent
    magnitude cut as in that work. However, the two estimates diverge
    when we use our absolute magnitude limits, since the Elbaz et al.
    (2007) method shows the redshift bias discussed above. Elbaz et al.
    (2007) use also volume densities to study the SFR-environment
    relation. However, we use here only the projected galaxy local
    density.  

\begin{figure}
   \centering
  \includegraphics[width=8cm]{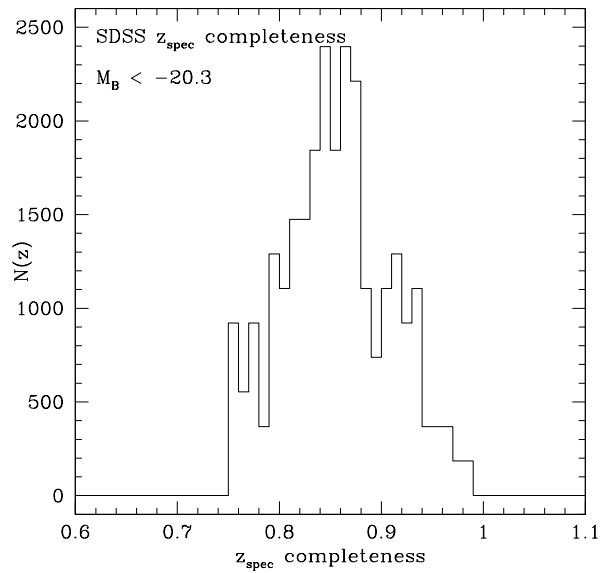}
\caption{Distribution of the completeness parameter calculated around each source in a SDSS subsample in a area of $10\times 10$ $\rm{deg}^{2}$ }
         \label{figura3b}
   \end{figure}

We compare our method also to the Nth nearest neighbor density
estimator (Dressler 1980; Capak et al. 2007; Guzzo et al. 2007,
Postman et al. 2005). According to this method, the projected density
around each galaxy is derived from the distance to the Nth neighbor,
r, which define a circular area whose surface density is $\Sigma =
N/{\pi}r^2$. We found that the two density estimators similarly trace
the same large scale structures.  In order to compare with similar
works based on accurate photometric redshifts, we use the accurate
GOODS-S phot-z to define a photometrically defined projected local
galaxy density. To compare to Feruglio et al. (2010) we use the 10th
nearest neighbor density estimator by adopting the same absolute
magnitude limit used above and a velocity interval equal to $3\times
{\sigma}_z(1+z)$, where ${\sigma}_z$ is the accuracy of the
photometric redshift (${\sigma}_z=0.01$ at $z < 1.1$ and $z_{AB} <
23$, similar to the one used in Feruglio et al. 2010). We find that
the two different definitions of the projected local galaxy density
agree with a very large scatter. The photometrically defined projected
local galaxy density can trace very high density regimes, however it
is not able to separate intermediate and low density regimes.

\begin{figure*}
\centering
\includegraphics[width=6cm]{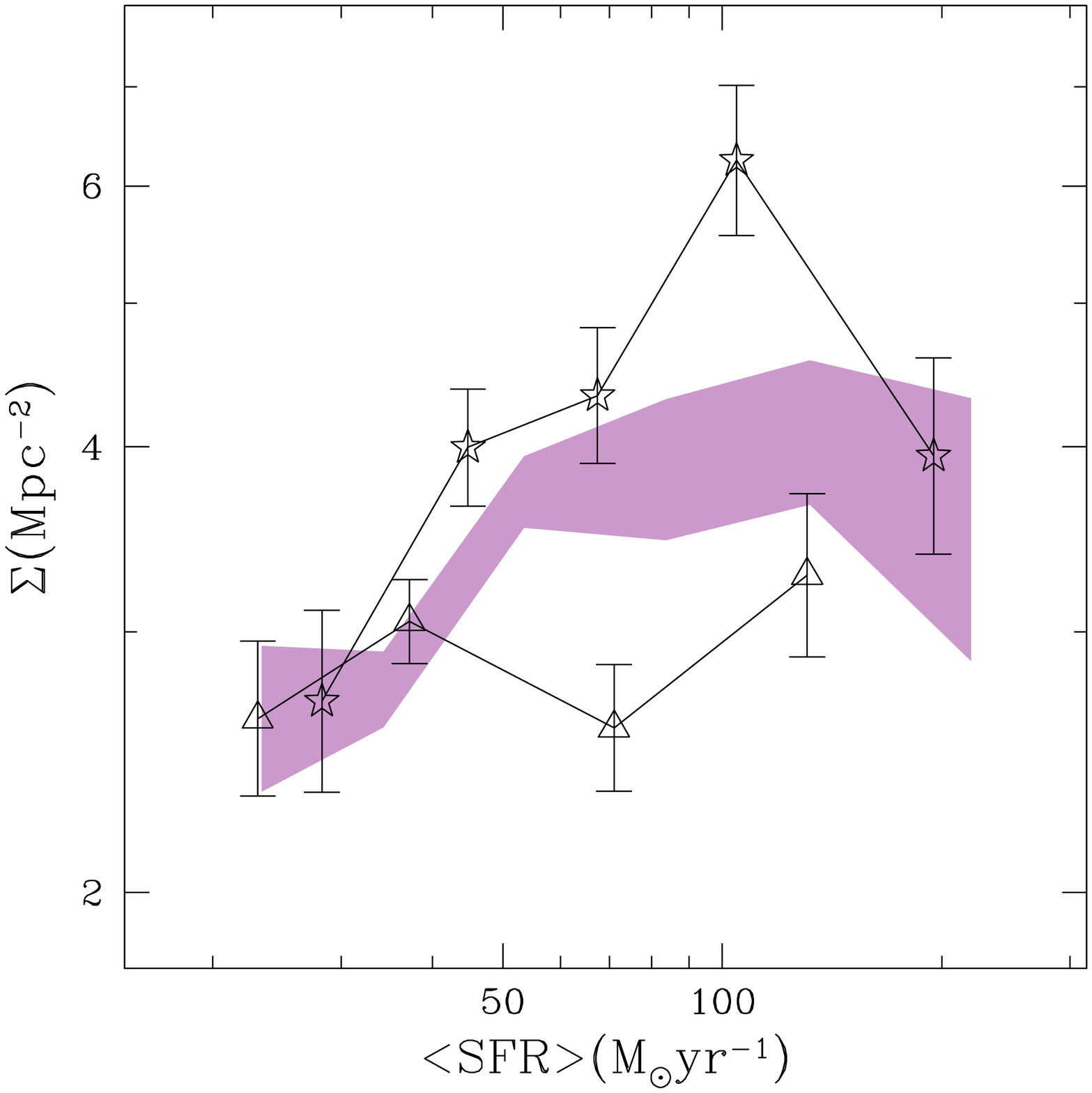}
\includegraphics[width=6cm]{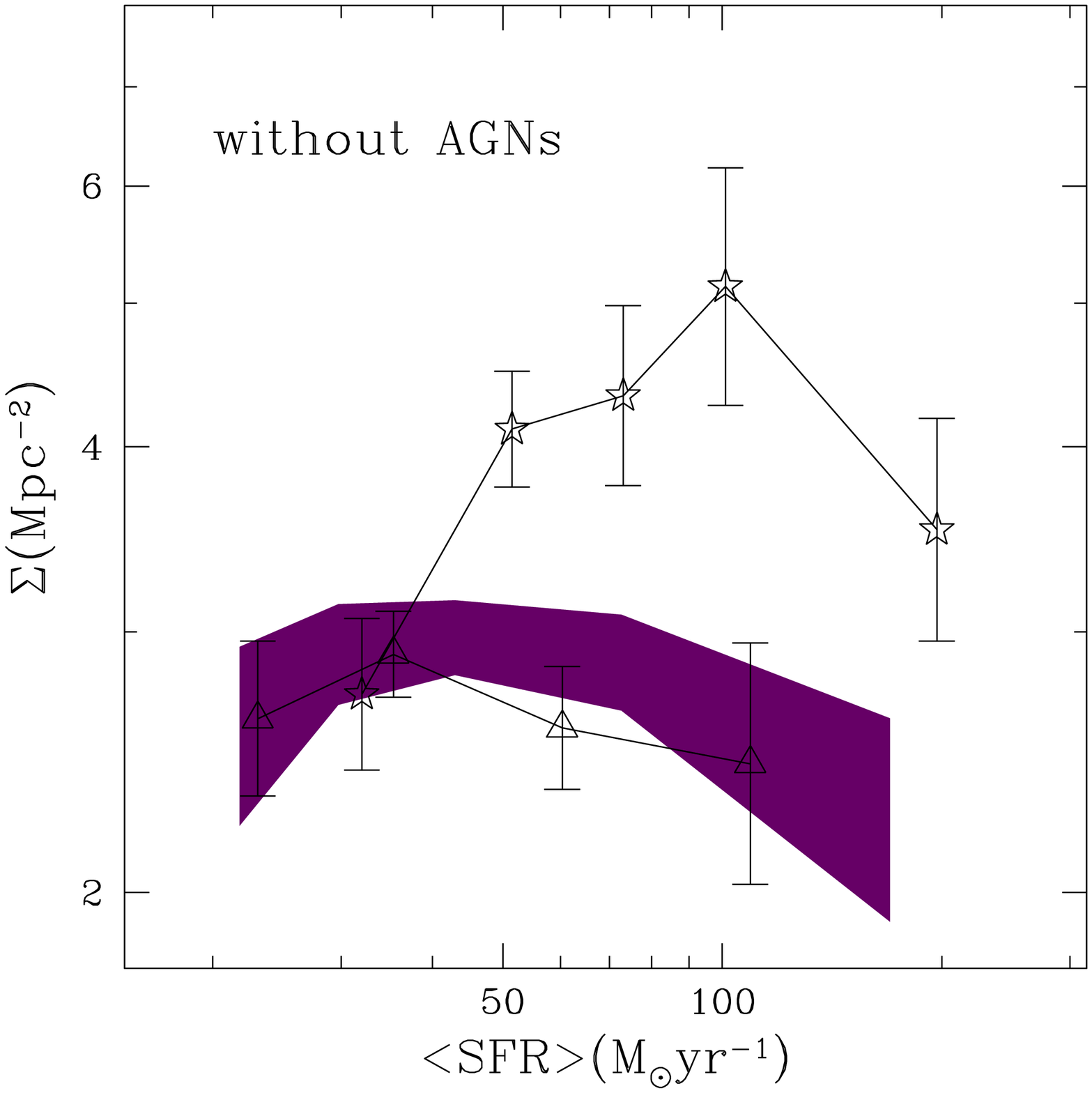}
\includegraphics[width=6cm]{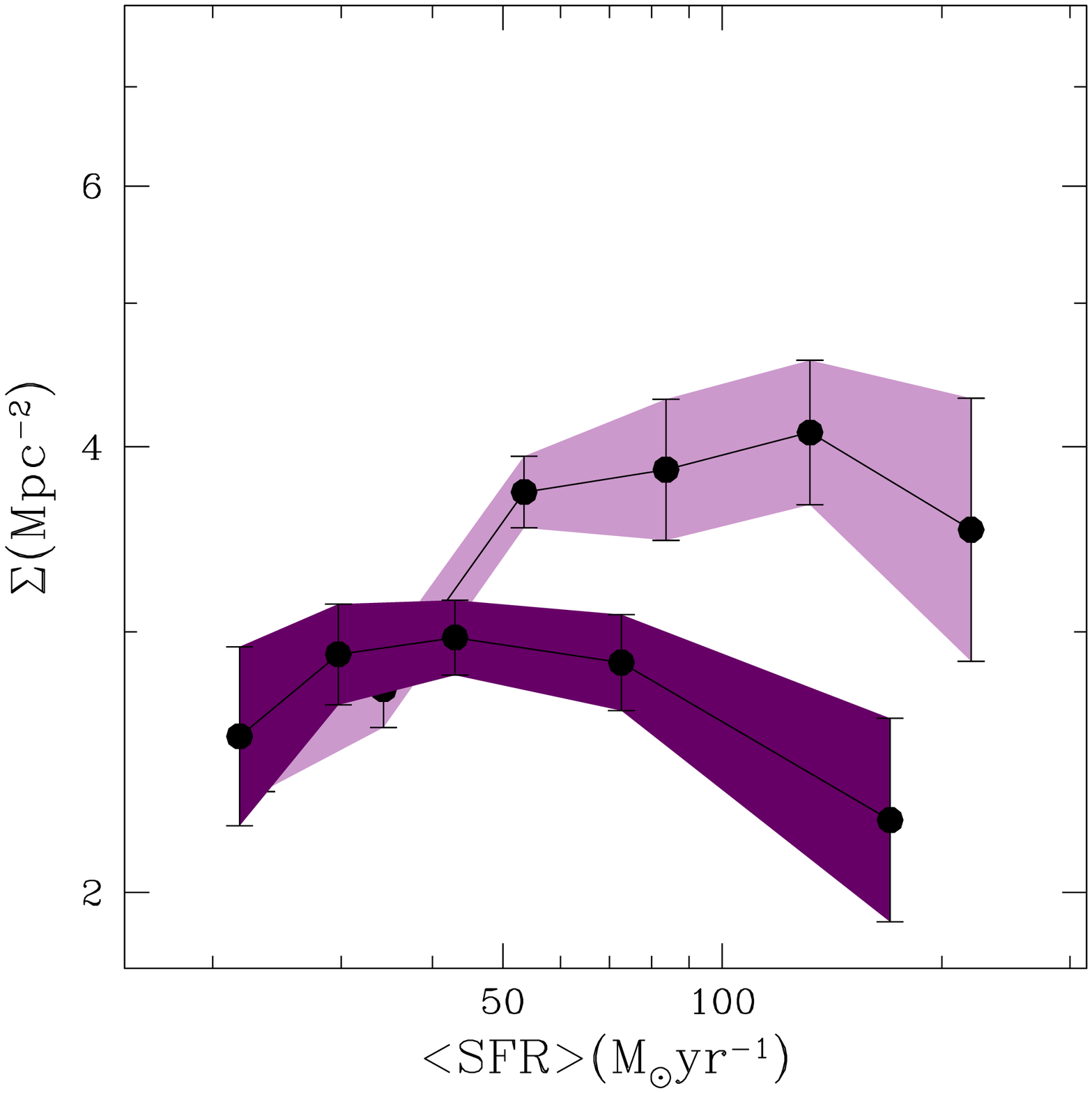}
\caption{Density-SFR relation. We show the relation obtained by including the low luminosity obscured AGNs (left panel) and the relation obtained after removing the AGNs from the sample (middle panel). In these panels the shaded regions show the global relations at $M/M_{\odot} > 8 \times 10^9$, empty triangles show the relation obtained in the low mass sample ($ 8 \times 10^9 < M/M_{\odot} < 5 \times 10^{10}$), stars show the relation obtained in the high mass sample ($M/M_{\odot} > 5 \times 10^{10}$).The right panel show the comparison of the global relations obtained by inclusion (pink area) and excluding (purple area) the AGNs in the sample.}
\label{figura4a}
\end{figure*}

We use a subsample of the whole SDSS DR7 galaxy spectroscopic sample described above, with spectroscopic redshift in the slice $0.01-0.1$ ($\sim$ 350000 galaxies), to build the density field around each star forming galaxy at low redshift.  In order to study the evolution of the density-SFR relation, we derive the projected local number density of the SDSS galaxies in the same way, within  $0.75$ Mpc and $\Delta \rm{v}=3000$ $\rm{km}{s}^{-1}$. Our absolute magnitude cut in the z band corresponds to $M_B=-20.75$ at redshift $\sim 1$. According to the B band rest frame luminosity function estimated at $0.75 < z < 1$ in Zucca et al. (2009), this cut is equal to $M^*+0.42$, where $M^*$ is the characteristic magnitude of the best-fit Schechter function. In order to sample the same luminosity range in the low redshift sample, we use as absolute magnitude cut $M_B=M^*(z=0.1)+0.42=-20.3$ mag, where $M^*(z=0.1)=-20.73$ mag is the characteristic magnitude of the Schechter function fitted to the rest frame B band LF at $ 0.1 < z< 0.35$ in Zucca et al. (2009). The B band is obtained from the SDSS  g band by using the $Kcorrect$ code of Blanton et al. (2007)

In order to obtain an estimate of the completeness correction for the SDSS star forming sample, we perform the same analysis applied to the high redshift sample to a subsample of SDSS galaxies in a region of $10 \times 10$ $\rm{deg}^{2}$, which includes several clusters at the redshift considered here. We use the mean value of the completeness level to correct the local galaxy density and we use the dispersion of the completeness distribution as error associated to this correction. As shown if Fig. \ref{figura3b}, the mean completeness level is 85\% $\pm 0.07$. As already shown in Popesso et al. (2005), the completeness level drops down to $\sim 70\%$ only in the very central region of the massive clusters within $\sim 0.1 Mpc$, due to the so called 'fiber-fiber collision' problem: fibers cannot be placed closer than 55 arcsec. Given the very large volume used for the estimate of the local galaxy density, the higher central incompleteness level does not affect significantly our density estimates.

\subsection{Stellar masses and Star Formation Rates}

Stellar masses and the total infrared luminosity ($L_{IR}$) of each
PACS source are derived according to the method described in
Rodighiero et al. (2010, and 2007), through a fitting of the whole
spectral energy distribution (SED) of each source. The stellar mass is
derived by fitting the UV-5.8 $\mu$m broad band photometric data with
the Bruzual \& Charlot (2003) stellar population synthesis models,
assuming a Salpeter IMF and a Calzetti extinction law, and by fixing
the redshift at the available spectroscopic redshift. For GOODS-S the
stellar masses are available from Santini et al. (2009) and are
calculated with the same stellar population synthesis models and
recipes (Rodighiero et al. 2010 and Santini et al. 2009 for further
details).

\begin{figure}
   \centering
\includegraphics[width=7cm]{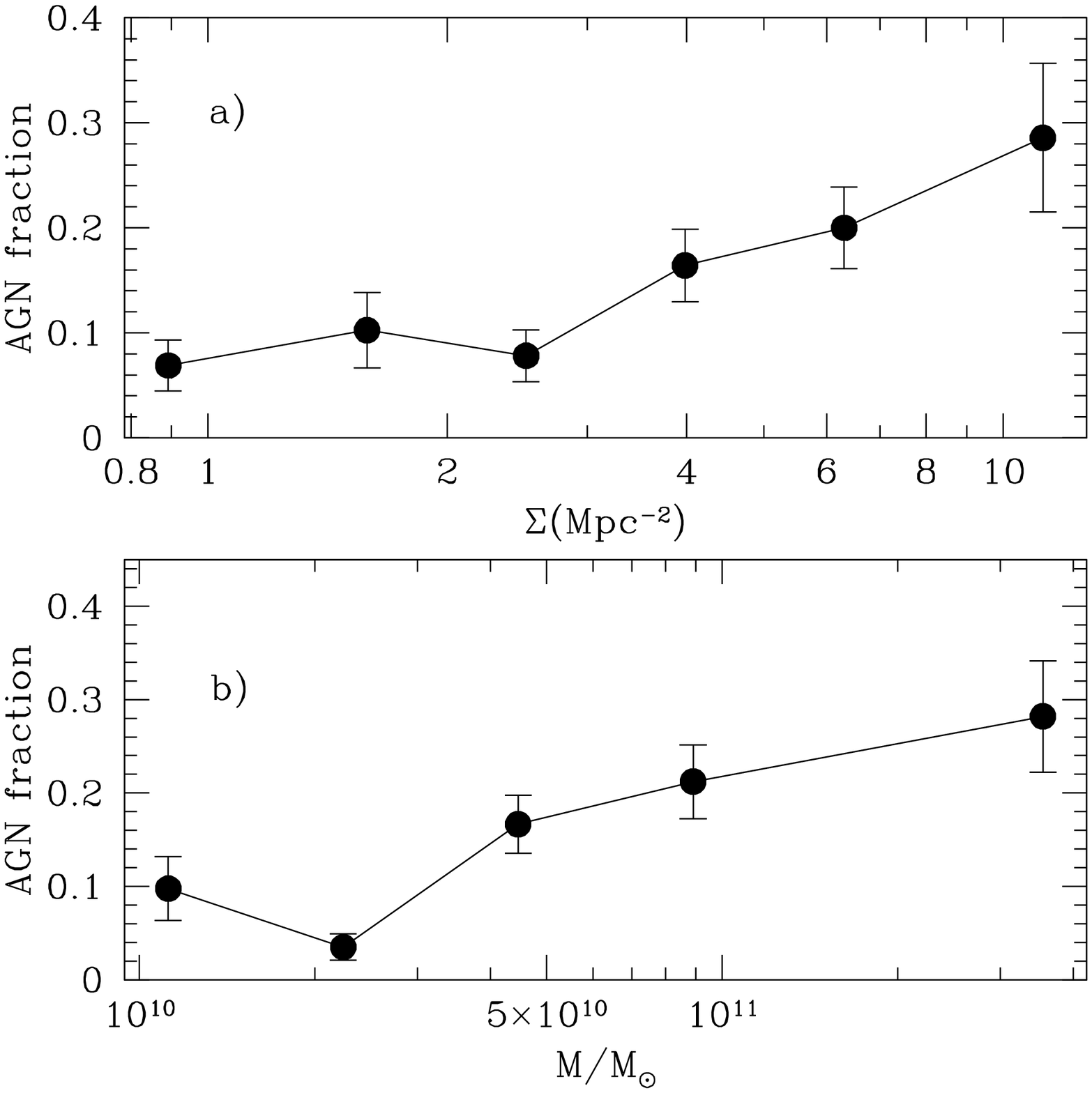}
\includegraphics[width=7cm]{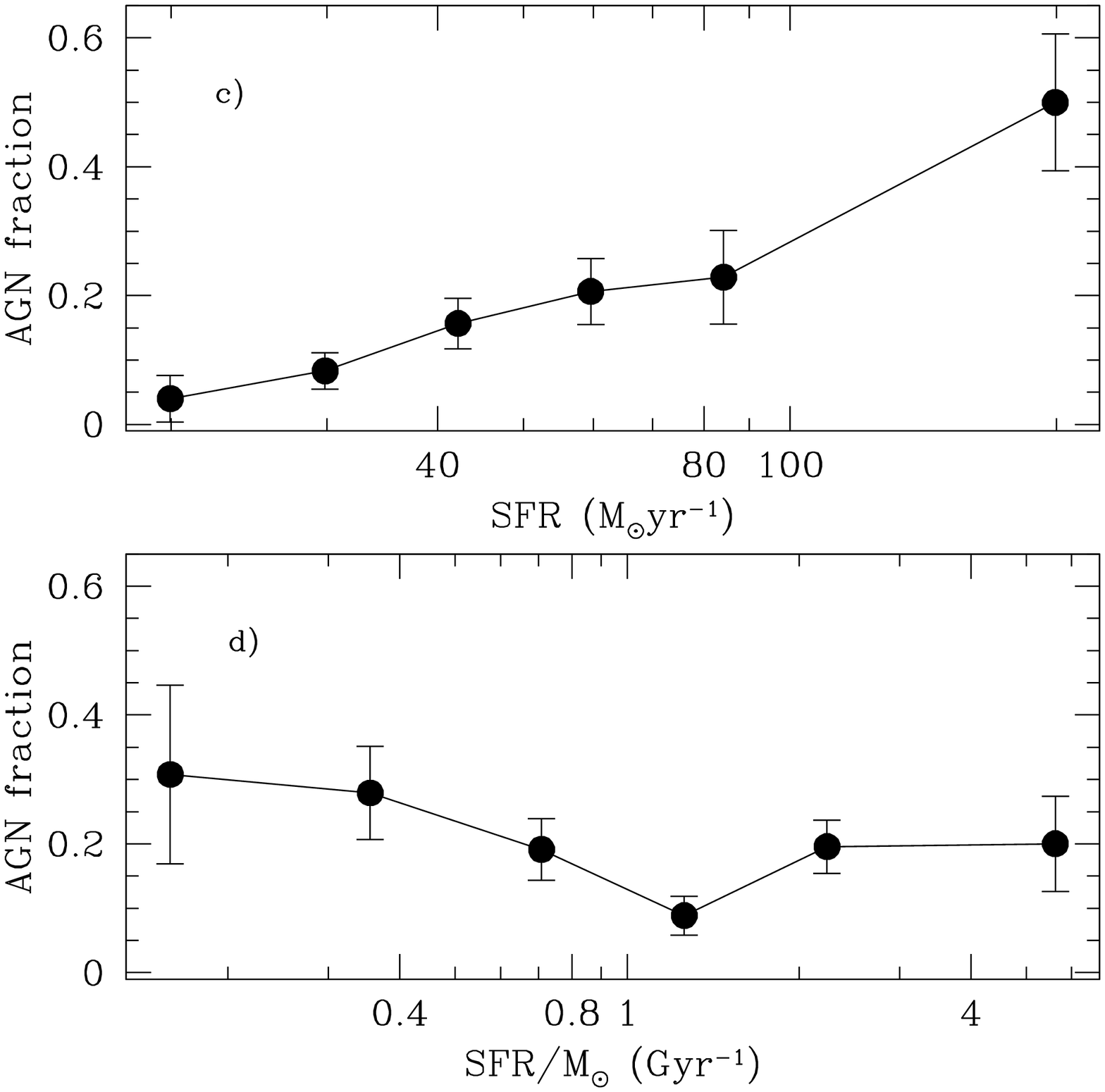}
      \caption{Fraction of AGN hosts in our sample of star forming galaxies as a function of density (panel a) and stellar mass (panel b), star formation rate (panel c) and specific star formation rate (panel d).}
         \label{figura5}
   \end{figure}

The $L_{IR}$ is derived in a similar way by fitting the optical-FIR
SED with a library of local templates, including the library from
Polletta et al. (2007) and few additional modified templates with
colder FIR emission (see Gruppioni et al. 2010 for more details) at a
given redshift. We compute $L_{IR}$ by integrating the best-fit SEDs
in the [8-1000]$\mu$m rest-frame range. The use of the whole SED
allows to obtain a more accurate and reliable estimate the K
correction (Rodighiero et al. 2010). For the sources with 24 $\mu$m
detection only, the fit is performed from optical up to MIR wavelength
only (hereafter R10 method).  For comparison with previous works we
calculate also $L_{IR}$ extrapolated from the 24 $\mu$m flux via the
Chary \& Elbaz (2001) method (hereafter CE01).  To check for
consistency, we compare the $L_{IR}$ estimate provided by the MIR
($L_{IR, 24 {\mu}m}$) only and the $L_{IR}$ based on the FIR fluxes
($L_{IR, PACS}$), for sources with both 24 $\mu$m and PACS detections.
Fig. \ref{figura2} shows the $L_{IR, 24 {\mu}m}/L_{IR, PACS}$ ratio
versus the $L_{IR, 24 {\mu}m}$ for the R10 and CE01 methods. In both
cases we see a positive correlation, confirmed by a Spearman test at
more than 7$\sigma$ level. There is a clear bias, whereby at lower
$L_{IR, 24 {\mu}m}$ the bolometric $L_{IR}$ is on average
underestimated with respect to the FIR based estimate. The CE01 method
produces a similar bias also versus the stellar mass as shown in the
right-hand panel of Fig. \ref{figura2}, which we do not find in the
R10 results. However, the R10 $L_{IR, 24 {\mu}m}/L_{IR, PACS} vs
L_{IR, 24 {\mu}m}$ relation is steeper and tighter, possibly
indicating a somewhat stronger bias. Thus, we use the CE01 $L_{IR, 24
  {\mu}m}$ as estimate of the $L_{IR}$ for the sources with only MIPS
detection. As already outlined in the previous section, the X-ray
detected AGNs follow the same trend as the normal star forming
galaxies. We perform a fit of the data with and without the AGNs and
we get consistent best fit lines.  Before combining PACS detected and
MIPS only detected sources, we correct the CE01 estimate $L_{IR, 24
  {\mu}m}$ by using the best fit obtained in the $L_{IR, 24
  {\mu}m}/L_{IR, PACS} vs L_{IR, 24 {\mu}m}$ relation.

The star formation rate is derived from the total infrared luminosity
$L_{IR}$ according to the Kennicutt law (1998). We checked that above
$L_{IR}/L_{\odot} > 10^{11}$, as in our selection, the contribution of
the uncorrected for extinction UV luminosity is negligible (on average
$\sim$ 3\%, much less than the mean 10 \% accuracy of our SFR
estimate).

\begin{figure}
   \centering
\includegraphics[width=8cm]{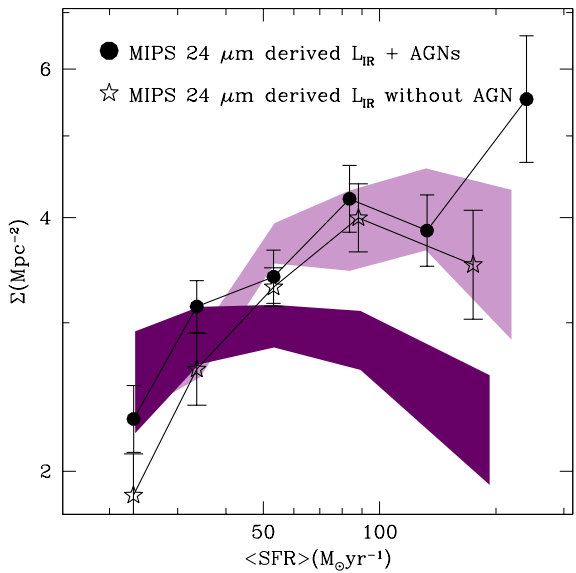}
     \caption{Density-SFR relation obtained using only  24 $\mu$m derived SFR. The filled points show the relation obtained with 24 $\mu$m only derived SFR with the AGN contribution, the empty stars show the same without AGN contribution. The positive correlation does not disappear after removing the AGN when 24 $\mu$m derived SFR are used. The shaded regions show the density-SFR relations estimated with PACS+MIPS data: pink region is the density-SFR relation with the AGN contribution, the violet region is the same without AGN contribution. }
         \label{confronto}
   \end{figure}

\section{Star formation activity and environment at z $\sim$ 1}

\subsection{The density-SFR relation}

We build the density-SFR relation to check which is the typical
environment of star forming galaxies at any given star formation rate. Thus, our
approach is to derive the mean density of the star forming galaxies per SFR bin.
This approach is less sensitive to selection effects due to
spectroscopic incompleteness. Indeed, the spectroscopic coverage of the
GOODS fields is spatially rather uniform, as pointed out in the
previous section. Instead, taking the average SFR per density bin is
more prone to biases because of the non uniform sampling of the
$L_{IR}$ luminosity function through the spectroscopic selection
function. We associate an error to the mean density per SFR bin through a jackknife technique. As done in Elbaz et at (2007), if n$_{\rm tot}$ is the number of galaxies  within a selected SFR interval, we randomly extract n$_{\rm jackknife}$ galaxies from this sample and computed their mean density. This process is repeated 500 times. The error bar on the typical  density for a given SFR bin is equal to the root mean square of the 500 measurements of the density divided by the square root of n$_{\rm tot}$/n$_{\rm jackknife}$. We checked that the error bar was robust by trying several values for N$_{\rm jackknife}$. In the highest SFR bins, where the number of galaxies gets small, we use the dispersion around mean density in the SFR bin divided by the square root of the number of galaxies in the bin. This technique is applied also to the study of the density-sSFR and sSFR-stellar mass relation in the following sections.

In order to disentangle the environment from the mass effect, we
perform this analysis in two bins of stellar masses with a cut at $5
\times 10^{10}$ $M_{\odot}$. The density-SFR relation is different in
the two mass bins. A 2-dimensional Kolmogorov-Smirnoff test shows that
the probability that the two distributions are drawn from the same
parent population is $1.6\times 10^{-4}$. A Spearman test performed on
the high mass sample gives a positive density-SFR correlation at the
3$\sigma$ confidence level (mean relation shown by stars in the left
panel of Fig. \ref{figura4a}). The same test applied to the low mass
sample, shows, instead, that SFR and density do not correlate (mean relation
shown by empty triangles in the left panel of Fig.  \ref{figura4a}).
In addition, as expected due to mass segregation, star forming
galaxies in the low mass bin lie in lower densities regimes with
respect to the systems of higher mass.

\begin{figure}
   \centering
  \includegraphics[width=8cm]{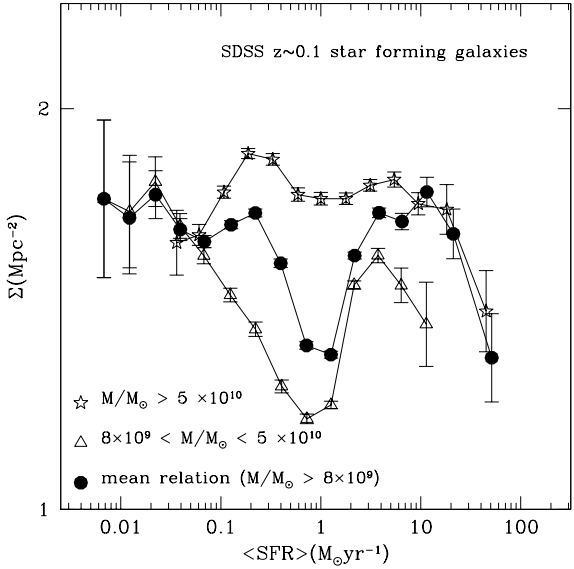}
\caption{Density-SFR relation in the local SDSS star forming sample. Filled points show the global relations at $M/M_{\odot} > 8 \times 10^9$, empty triangles show the relation obtained in the low mass sample ($ 8 \times 10^9 < M/M_{\odot} < 5 \times 10^{10}$), stars show the relation obtained in the high mass sample ($M/M_{\odot} > 5 \times 10^{10}$).}
         \label{sloan1}
   \end{figure}

\begin{figure*}
\centering
\includegraphics[width=6cm]{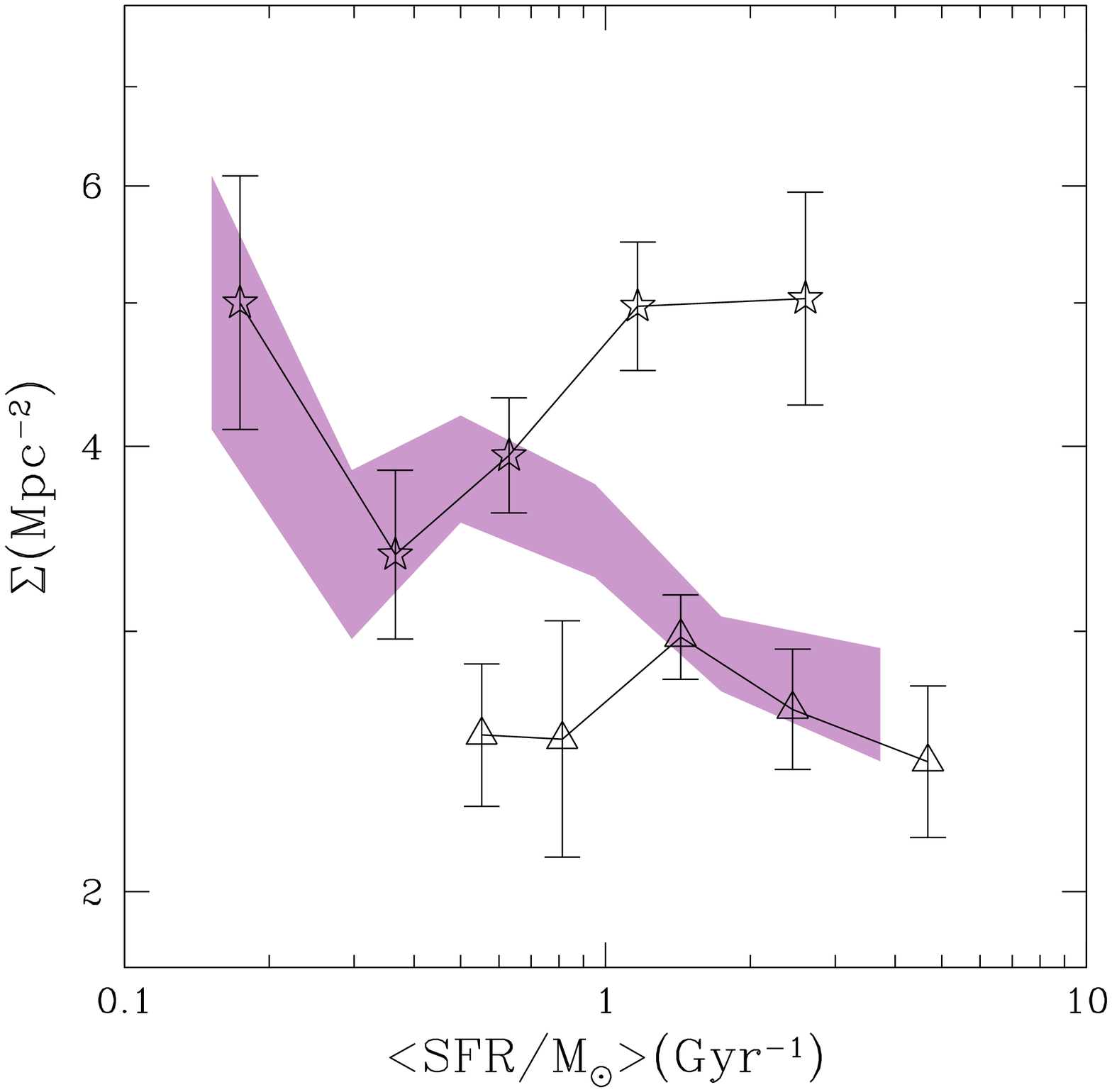}
\includegraphics[width=6cm]{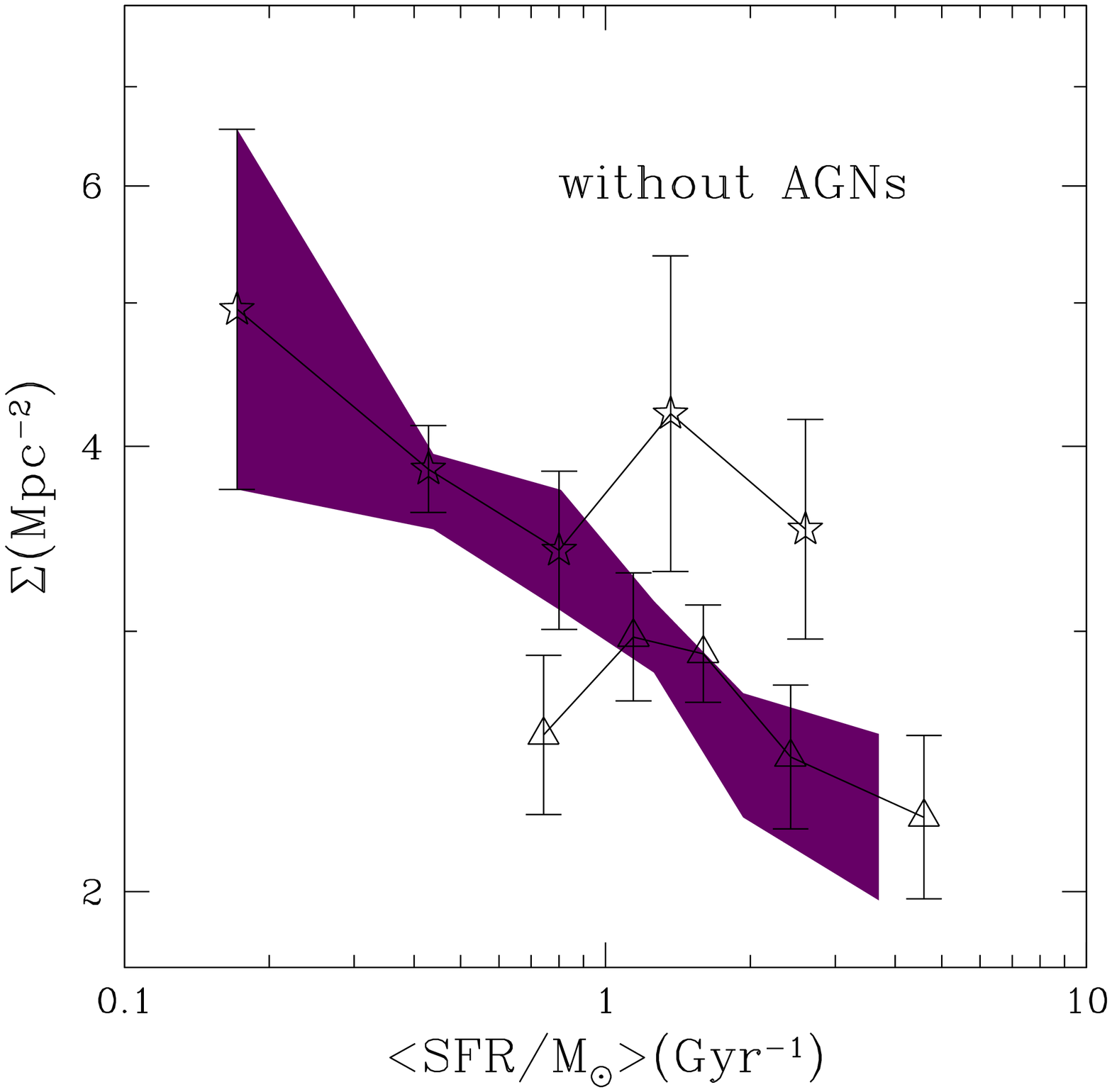}
\includegraphics[width=6cm]{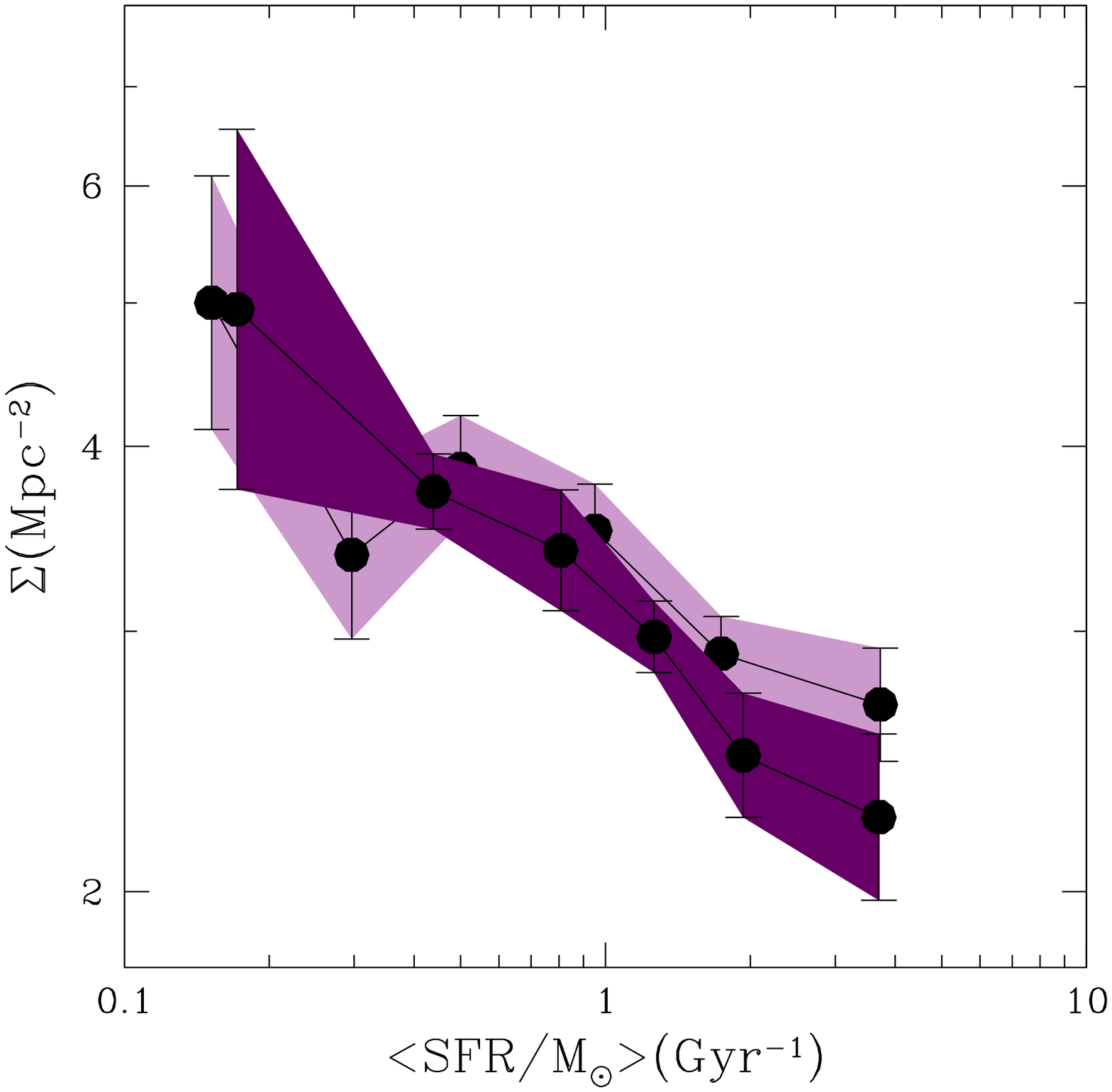}
\caption{Density-sSFR relation. We show the relation obtained by including the low luminosity obscured AGNs (left panel) and the relation obtained after removing the AGNs from the sample (middle panel). In these panels the shaded regions show the global relations at $M/M_{\odot} > 8 \times 10^9$, empty triangles show the relation obtained in the low mass sample ($ 8 \times 10^9 < M/M_{\odot} < 5 \times 10^{10}$), stars show the relation obtained in the high mass sample ($M/M_{\odot} > 5 \times 10^{10}$).The right panel shows the comparison of the global relations obtained by inclusion (pink area) and excluding (purple area) the AGNs in the sample.}
\label{figura4b}
\end{figure*}

To check if the positive
    correlation observed in the high mass bin is due to a
    residual mass segregation effect, we check through a KS test that
    the mass distribution is the same in all SFR bins. According to
    the KS statistics the probability that the mass distributions of
    galaxies in different SFR bins are drawn from the same parent
    distribution is 85\% in the worst case. We  conclude that the
    the rising of the density-SFR relation in the high mass bin is due
    to an environmental effect. The density-SFR relation estimated in
    the whole sample gives a 3.5$\sigma$ positive correlation (shaded
    pink area in Fig. \ref{figura4a}). We ascribe the correlation to the fact
    that massive galaxies are dominating at high star formation
    rates.

Before comparing the redshift $\sim 1$ density-SFR relation with the local reference, we have first to remove the AGN from the sample. Indeed, we exclude the AGN from the low redshift SDSS star forming sample since the AGN contribution to the $H\alpha$ emission can lead to a strong overestimation of the star formation rate. 
Once the AGN host galaxies are removed from the sample, the observed
correlations become less significant and flatter, as shown in the
middle panel of Fig. \ref{figura4a}. The Spearman test gives a
much higher probability of no correlation both in the high mass bin
and in the global relation ($\sim$ 20\%).  We also point out that the AGN with only MIPS detection lie all at low luminosities (the first three SFR bins in the density-SFR relation).  Their removal do not affect at all the relation, meaning that even if the 12 $\mu$m rest frame flux can be biased by the AGN contribution towards higher $L_{IR}$, this bias is not affecting our relation.

The effect due to the AGN
removal is to remove massive galaxies at high density. Indeed, the AGN
fraction in our sample is a rising function of density and mass and
star formation rate (Fig. \ref{figura5}). This result is in agreement
with previous results obtained at low redshift and based on obscured
AGNs (Kauffmann et al. 2003) and radio-loud (Best et al. 2005) AGN in
the SDSS. More recently Silverman et al. (2009) find in the COSMOS
field a similar trend at z $\sim$ 0.5, but with a fraction of AGN 5
times smaller. The discrepancy is likely due to to the fact that the
deep 2Ms Chandra observations of the GOODS fields allow to sample a
larger portion of the $L_X$ luminosity function at the steep faint-end
with respect to the shallower XMM survey of the COSMOS field.

In order to test whether removing massive galaxies irrespective of the
AGN presence can destroy the density-SFR correlation in the same way,
we performed a MonteCarlo simulation by randomly removing a million
times massive galaxies from our sample in the same number of the AGNs
and performing the Spearman test on the remaining galaxies. Only in
0.5\% of the cases removing massive galaxies leads to the same or
lower probability of correlation as removing the AGN. The probability
is somewhat higher (1.5\%) is we remove randomly chosen galaxies
irrespective of the mass. As additional test we removed for a
    million times randomly chosen galaxies in the same number and with
    the same mass distribution of the AGN in our sample. Similarly, we
    find that in 1.6\% of the cases, the Spearman test gives the same
    or lower probability of correlation as removing the AGN. In
addition, the higher the AGN fraction left in the sample, the higher
the correlation probability.  Thus we consider AGN host galaxies as a
good candidate to cause the reversal of the density-SFR relation in
our sample.

Once the AGN hosts are removed from the high redshift star forming
    galaxy sample as in the low redshift sample, the agreement between
    the behavior of the density-SFR relation at $z\sim 1$ and $z\sim
    0.1$ is remarkable. Indeed, as shown in Fig. \ref{sloan1}, although
    in a much larger SFR range, we do not observe any correlation in
    the SDSS star forming sample as confirmed by the Spearman test.
    The mean relation is shown by filled points in Fig. \ref{sloan1}.
    Once we split the SDSS sample in two mass bins as we do for the
    $z\sim 1$ galaxies, we observe the same mass segregation effect,
    e.g. the higher the mass, the higher the mean density. We do not
    observe any correlation in any mass bin similarly to the $z\sim 1$
    galaxies without AGN hosts. We do not compare here the absolute
    value of the local density of star forming galaxies at $z \sim 1$
    and $z \sim 0.1$, because the density parameter is not estimated
    exactly in the same comoving volume.

The comparison of our density-SFR relation with analogous results in the literature is not straightforward. Indeed, all previous works use a different approach to derive the same relation: they estimate the mean SFR per density bin. This is equivalent to study the galaxy type mix per density regime rather than investigate which is the typical environment of galaxy with a given instantaneous star formation rate.  However, qualitatively we find a good agreement with Elbaz et al.  (2007) and Cooper et al. (2008) if AGN hosts are kept in our analysis.  The density-SFR relation obtained after removing the AGN contribution is much flatter than in Elbaz et al. (2007). Indeed, as shown in Fig.
    \ref{confronto}, when we use 24 $\mu$m only derived SFR, the
    density-SFR relation is steeper than the one based on PACS data if
    AGN are kept in the sample. In addition, the removal of AGN from
    the sample does not destroy the correlation as observed in the
    relation obtained with PACS derived SFR. This is confirmed by a
    Spearman test, which gives a positive correlation at $3.5\sigma$
    level even after removing the AGN.  This can be explained by the
    bias observed in the 24 $\mu$m only derived $L_{IR}$ in Fig.
    \ref{figura2}.  The same comparison is not possible with Cooper
et al. (2008) because they cannot identify and remove AGN in the DEEP2
sample.  We find a better agreement with Feruglio et al. (2010) who
exclude AGN from the analysis and find a flat relation.

\begin{figure}
   \centering
  \includegraphics[width=8cm]{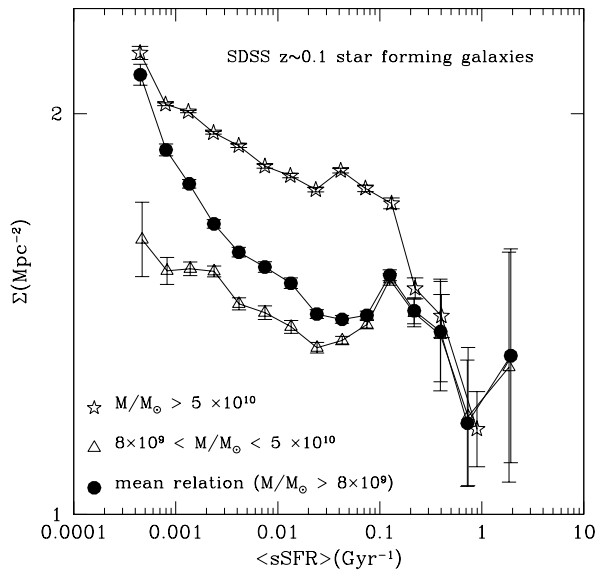}
\caption{Density-sSFR relation in the local SDSS star forming sample. Filled points show the global relations at $M/M_{\odot} > 8 \times 10^9$, empty triangles show the relation obtained in the low mass sample ($ 8 \times 10^9 < M/M_{\odot} < 5 \times 10^{10}$), stars show the relation obtained in the high mass sample ($M/M_{\odot} > 5 \times 10^{10}$).}
         \label{sloan2}
   \end{figure}

\subsection{The density-sSFR relation}

We analyze here the density-specific SFR (sSFR=SFR/$M_{\odot}$) 
relation. Due to the known anti-correlation between the sSFR and the
galaxy stellar mass (Elbaz et al. 2007, Rodighiero et al. 2010),
binning in stellar mass implies a selection in sSFR. We show the
result also for the low and high mass samples (same stellar mass cut
as in the previous analysis) separately, to analyse the different
contributions to the global density-sSFR relation.  The mean relation
is shown in the left panel of Fig. \ref{figura4b} (shaded region).
The Spearman test reveals a negative correlation with probability of
no correlation $P(r_S)=0.23\%$.  The analysis of the contributions
from the different mass samples reveals that: a) in the high mass bin 
sSFR and density marginally correlate (90\% confidence level according
a the Spearman test, empty stars in left panel of Fig.
\ref{figura4b}), as already found in the same mass bin by Elbaz et al.
(2007); b) at lower stellar masses there is no correlation (empty
triangles in left panel of Fig. \ref{figura4b}); c) due to mass
segregation, high mass systems lie generally in dense regions while
low mass systems prefer low density environments; d) the high mass
systems dominate the low sSFR end; e) the low mass systems dominate
the high sSFR bins. This mass segregation leads to a global
density-sSFR anti-correlation.

The removal of AGN hosts from the sample does not affect significantly
the result. Indeed, as shown in the panel d) of Fig. \ref{figura5},
the AGN fraction is relatively constant at any sSFR. Thus the effect of
removing AGNs is to reduce the number of galaxies populating any sSFR
bin and consequently to enlarge the error bars. This directly implies
that the level of star formation activity of AGN hosts is consistent
with the star formation level of star forming galaxies at the same
mass.

The anti-correlation observed in the star forming sample at $z
    \sim 1$ is consistent with the behavior observed at $ z \sim 0$,
    as shown in Fig. \ref{sloan2}. The local star forming sample show
    the same mean anticorrelation (filled points in Fig. \ref{sloan2})
    as confirmed by the Spearman test. Differently from the high
    redshift case, we observe a significant anti-correlation also in
    the high (stars) and low mass bins (triangles). However, this can
    still be due to mass segregation since a Spearman test reveals a
    very significant ( $ > 15\sigma$) positive correlation between
    mass and local density in the individual mass bins.

For the same reasons described in the previous section, the comparison
of our relations with the previous work is not straightforward.
Nonetheless, as already outlined, we can reproduce the marginal
positive correlation observed by Elbaz et al. (2007) between sSFR and
density in the high mass bin. We ascribe this marginal correlation to
the AGN contribution. We find good agreement with Cooper et al. (2008)
who observe a global density-sSFR anti-correlation. We find partial
agreement with Feruglio et al. (2010) who do not find any relation in
fixed stellar mass bins. However, they do not provide a definitive
result about the combination of the contributions of the different
mass bins to the global sSFR-mass relation.


   \begin{figure*}
   \centering
\includegraphics[width=6cm]{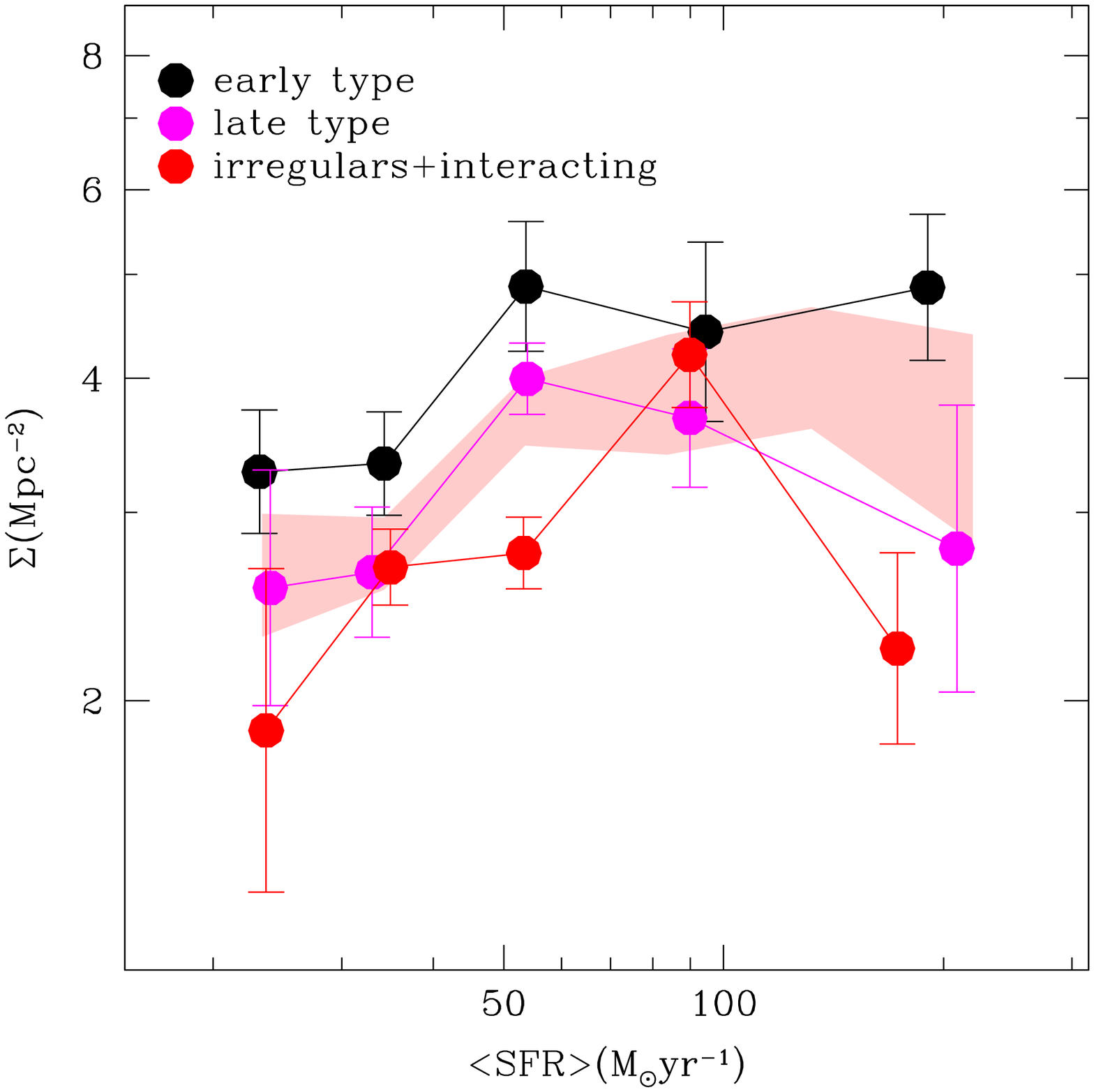}
\includegraphics[width=6cm]{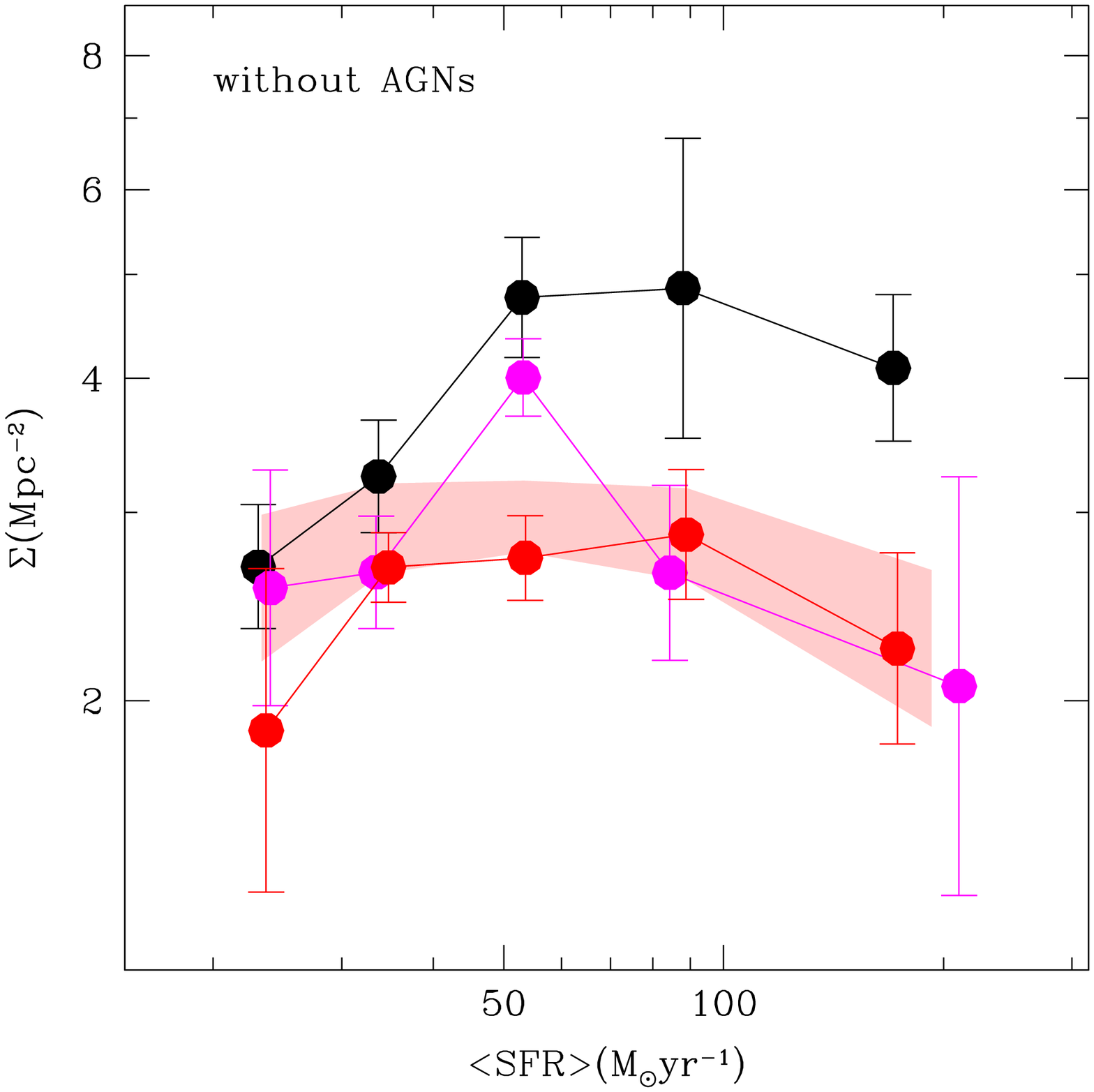}
\includegraphics[width=6cm]{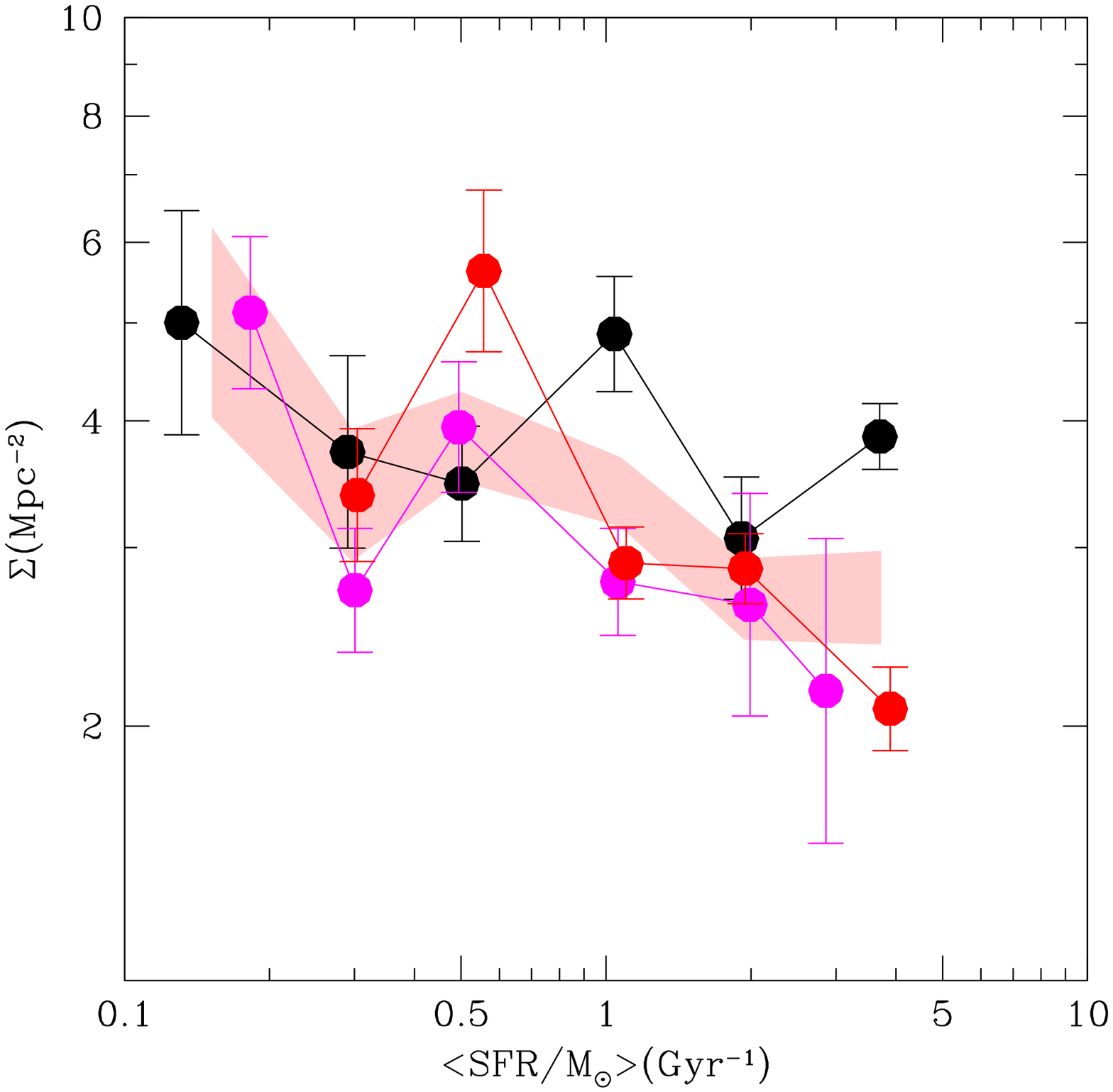}
\caption{The left panel shows the density-SFR relation per morphological type. Black points refer to early type galaxies, magenta points refer to late type galaxies, and red points to galaxies with disturbed morphology. The central panel shows the same relation without the contribution of AGN hosts. The left panel shows the density-sSFR relation per morphological type. The symbols have the same meaning in all panels. The shaded regions in all panels shows the global trend of the considered relation.}
         \label{figura7}
   \end{figure*}

\subsection{The morphological analysis}

In this section we analyse the previous relation per morphological
type in order to link the galaxy SFR (sSFR) and environment to the
galaxy morphology. The morphological classification is performed by
visual inspection of all objects in our sample. We identify 5
morphological types: 49 spheroidal, 71 bulge dominated objects, 73
disk dominated spirals, 97 irregulars, 34 interacting galaxies. We
exclude two objects that turn out to be stars. In order to have enough
statistics, we group the spheroidals and the bulge dominated galaxies
as early type galaxies, and the irregulars and interacting objects as
galaxies with disturbed morphology. Thus, early type galaxies account
for 37\% of the sample, late type galaxies for 22\% and galaxies with
disturbed morphology for the remaining 41\%.  Fig. \ref{figura7} shows
the density-SFR relation (left panel) for the different morphological
types. According to the Spearman test none of the three considered
morphological types shows a significant correlation. The mean density
slightly increases from the late to the early types. Below 50
$M_{\odot}yr^{-1}$ the late type and irregular galaxies are
dominating. As a consequence  the mean density is low around $2.6\pm0.4$
$\rm{Mpc}^{-1} $. Above that threshold the galaxies are almost equally
distributed in the three morphological types. This brings the mean density
at $3.9\pm0.5$ $\rm{Mpc}^{-1}$ above 50
$M_{\odot}yr^{-1}$ . This shows that the morphology-density
relation is already in place among LIRGs at redshift $\sim$ 1 and how
the morphological type mix relates to the reversal of the density-SFR
relation. As for the global relation, the removal of the AGN hosts
from the sample leads to a flattening of the density-SFR relation for
any morphological type, as shown in the central panel of Fig.
\ref{figura7}. A Spearman test confirms that there is no correlation for all morphological types. This is
also confirmed by the large error bars in the figure.

The right panel of Fig. \ref{figura7} shows the density-sSFR relation per morphological type. All three galaxy types follow the global relation. We do not show the effect of the AGN removal because, as for the global relation, the  trend remains unchanged while the error bars increase due to the lower statistics.

\begin{figure*}
   \centering
\includegraphics[width=8cm]{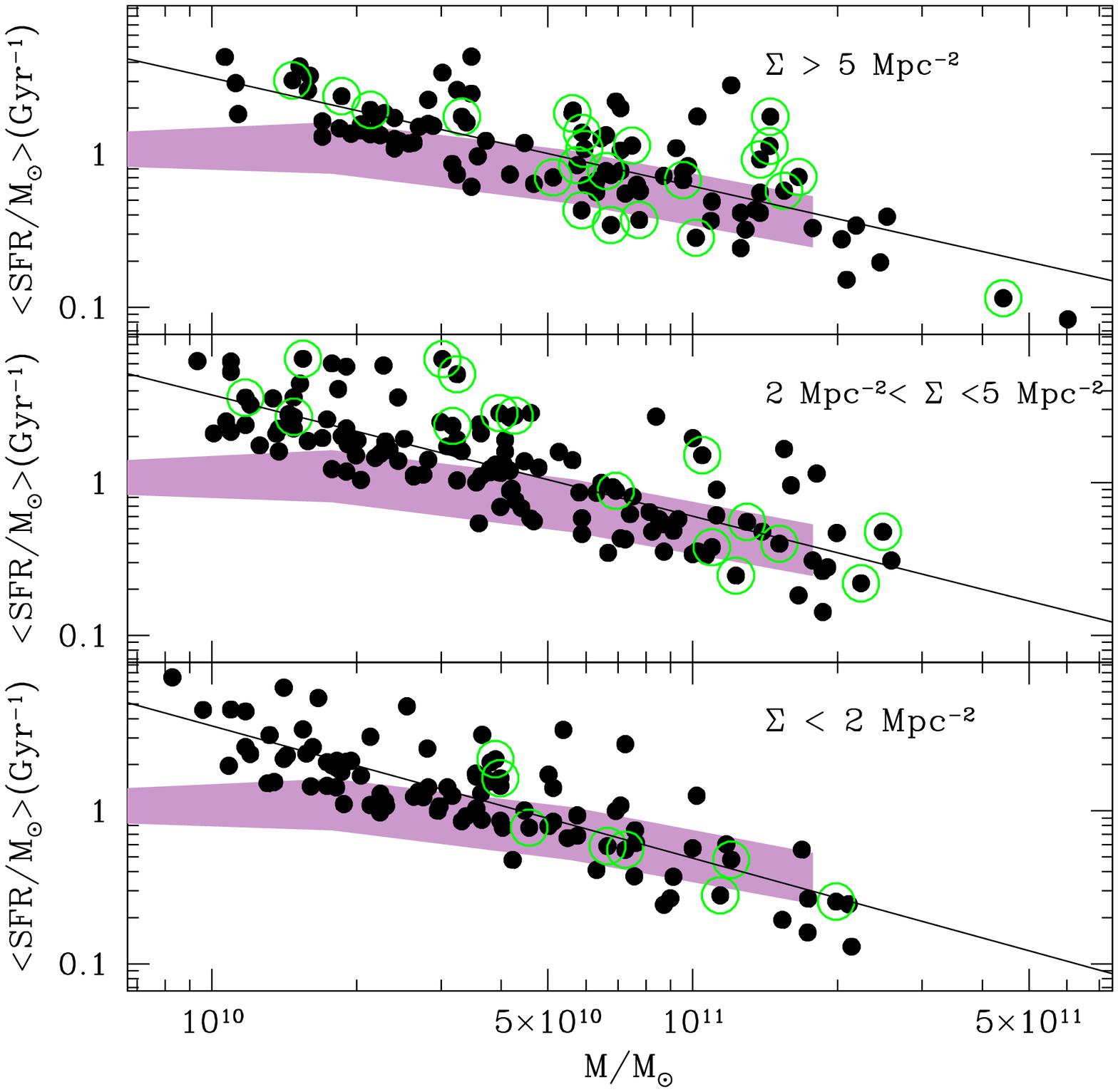}
\includegraphics[width=8cm]{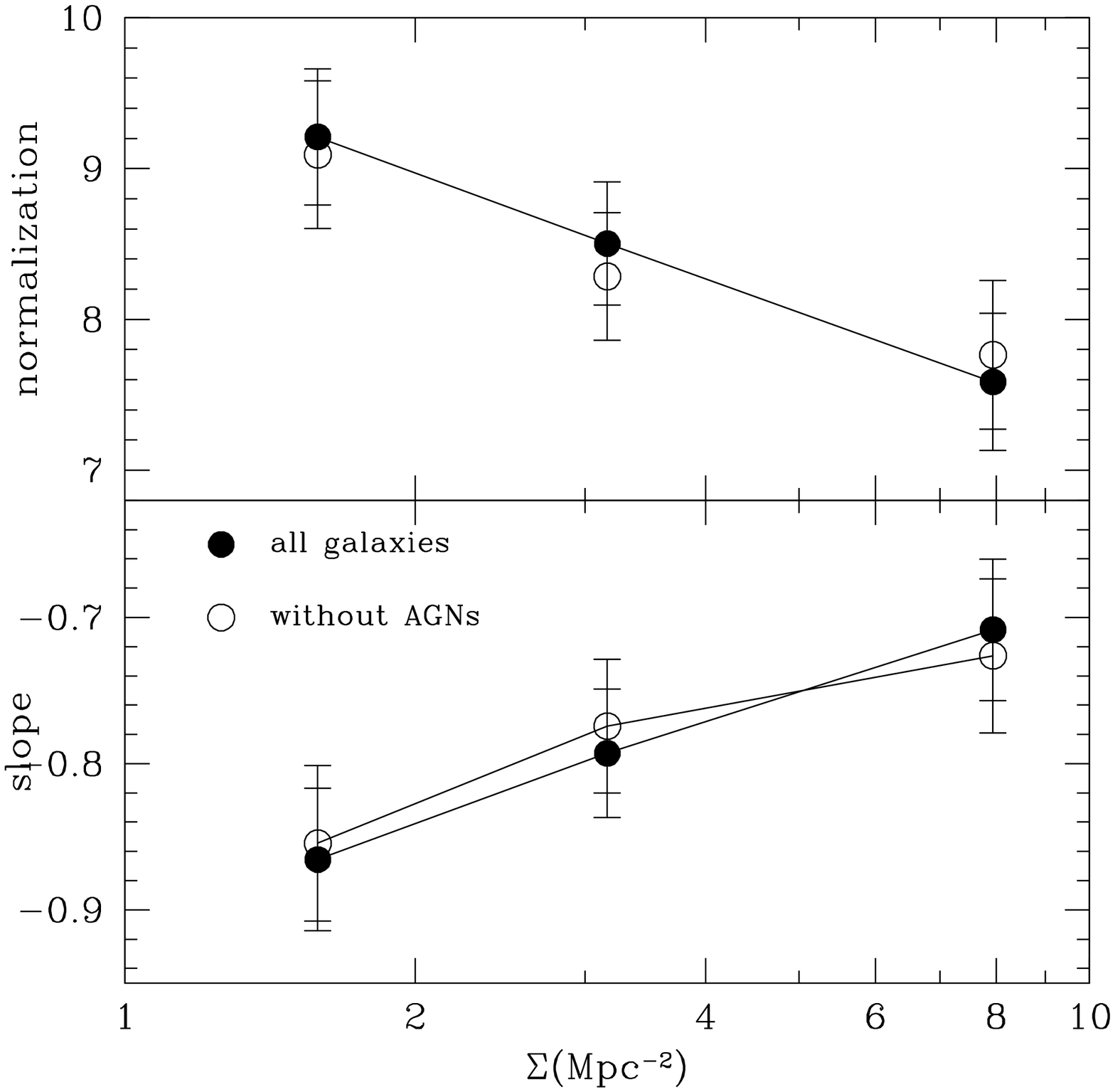}
      \caption{ sSFR-stellar mass relation in three density bins (left panel). The green circles identify AGN hosts. The solid line shows the best fit to the relation. The shaded purple region is the relation obtained via stacking in Rodighiero et al. (2010) based on a IRAC selected sample. The right panel shows the best fit parameters (filled points, normalization in the top panel, slope in the bottom panel) as a function of the bin mean density. The empty points show the best fit parameters obtained after excluding the AGN hosts.}
         \label{figura6}
   \end{figure*}

\subsection{The sSFR-stellar mass relation}

\begin{figure}
   \centering
  \includegraphics[width=8cm]{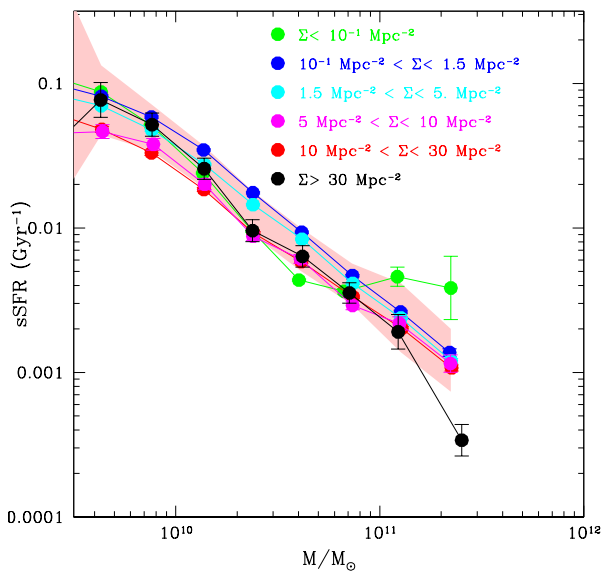}
\caption{sSFR-stellar mass relation per density bin in the local SDSS star forming sample. The shaded region is the mean relation based on the full star forming sample. The relations are color-coded as a function of density as explained in the figure.}
         \label{sloan3}
   \end{figure}

In order to gain more insight into the relation between galaxy SFR and
environmental conditions we study also the sSFR- stellar mass relation
in three density bins, to isolate low, intermediate and high density
regimes. Each bin contains at least 100 sources.  In order to control
the selection effects, we compare our relation with the relation of
Rodighiero et al. (2010) based on the stacking of an IRAC selected
sample. As shown in Fig. \ref{figura6} the R010 relation (left panel,
shaded region) is consistent with our relation. As expected, we
slightly overestimate the sSFR-mass relation at the low mass end due
to our $10^{11}$ $L/L_{\odot}$ luminosity cut. We find a rather
significant difference ( at the $3\sigma$ level) rising from low to
high density regimes.  The discrepancy is confirmed by a
    2-dimensional Kolmogorov-Smirnoff test showing that the mutual
    sSFR and stellar mass distributions have a probability of $6\times
    10^{-3}$ to be drawn from the same parent distribution from the
    lowest to the highest density bin. Partly this is due to mass
    segregation. Indeed a unidimensional KS test applied to the
    stellar mass distribution between the lowest and highest density
    bins, shows with high significance that they are not drawn from
    the same parent distribution. The same test applied to the sSFR
    gives a somewhat higher probability, ~20\%. The left panel of Fig.
\ref{figura6} shows the sSFR-mass relation in the three density bin.
We fit the best fit line by linear regression in the log-log
    space and show the best fit parameters as a function of the mean
density in the central panel. The errors are estimated through a
jackknife technique. We see a clear trend of the relation which is
flattening towards higher local density.  The removal of AGN hosts
does not change the trend. This is due to the fact that the AGN
fraction is equally distributed at all sSFR (panel d of Fig.
\ref{figura5}). So removing the AGN hosts mainly enlarges the error
bars due to the lower statistics. Fig. \ref{sloan3} shows the
same analysis applied to the SDSS local star forming sample. We observe a
    change in the sSFR-mass relation as a function of the local
    density.  This was already pointed out in Kauffmann et al. (2004), although they consider all SDSS galaxies and not only star forming systems as in this work. The effect is visible only at the high mass end of the
    relation, $ M/M_{\odot} > 10^{11}$, whereby the higher the density
    the lower the mean sSFR of the star forming galaxies. We point out
    that Peng et al. (2010) studies the same relation using SDSS
    galaxies without reporting any environmental effect. However, we
    also point out that Peng et al. (2010) consider a much larger range of
    stellar masses, from $10^7$ to $10^{12}$ $M/M_{\odot}$. As shown
    in Fig. \ref{sloan3}, below $10^{11}$ $M/M_{\odot}$, we do not
    observe a deviation as a function of density with respect to the
    mean relation (shaded region). Although they do not dominate in
    mass, low mass galaxies surely dominate in number with respect to
    massive galaxies and determine statistically the slope of the
    relation. Thus, a linear fit to the sSFR-mass relation in the
    log-log space, as done in Peng et al. (2010), provides slopes
    consistent in any density regime.

\begin{figure*}
\centering
\includegraphics[width=8cm]{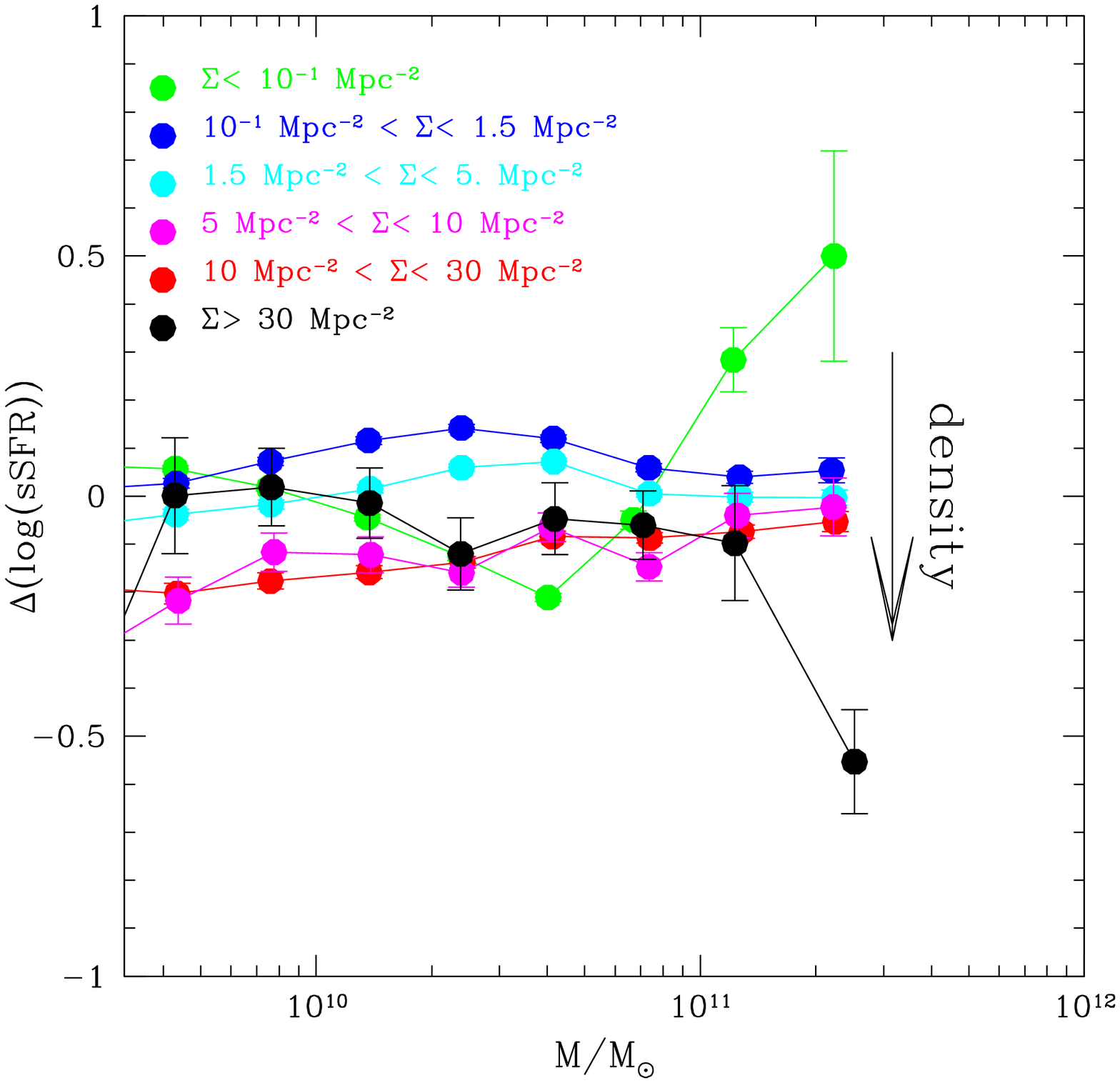}
\includegraphics[width=8cm]{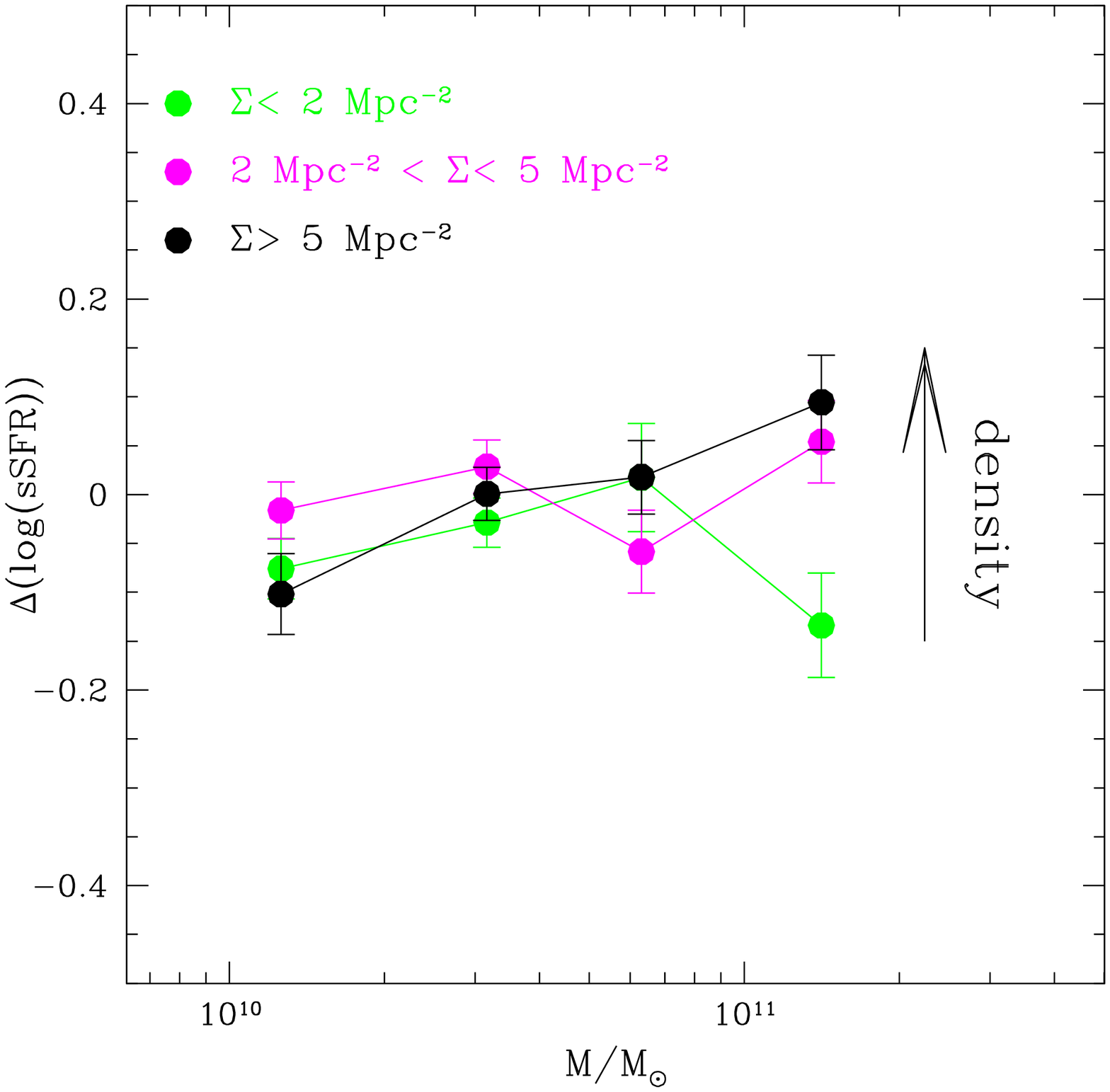}
\caption{Residual $\Delta(sSFR)$ as a function of mass and density
      with respect to the mean sSFR-mass relation of the local SDSS
      star forming sample (left panel) and of the $z\sim1$ star
      forming sample (right panel). The points are color-coded as a
      function of local density as explained in the panels.}
\label{sloan4}
\end{figure*}

\begin{figure}
   \centering
  \includegraphics[width=8cm]{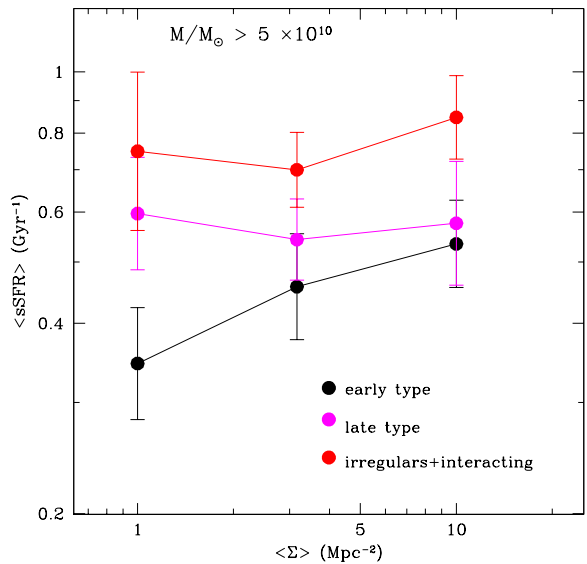}
\caption{Mean sSFR per density bin in the high mass bin ($M/M_{\odot} > 5 \times 10^{10}$) and per morphological type. The color coding as function of morphology is shown in the figure.}
         \label{early}
   \end{figure}

Fig. \ref{sloan4} shows the direct comparison of the density
       dependence of the sSFR-mass relation at low and high redshift.
       The figure shows the residuals $\Delta(log(sSFR))$ of the
       sSFR-mass relation per density bin with respect to the mean
       relation in the local SDSS star forming sample (left panel) and
       in the high redshift star forming sample (right panel). The
       deviation from the mean relation at the high mass end is clear
       at $z \sim 0.1$ (left panel) and,  although much less significant
       ($2.5\sigma$ effect), still visible in the $z\sim 1$ case (right panel).  As
       indicated by the arrows, we find an indication that the effect is going in opposite direction at high redshift with respect to the local analog.
       To explain this reversal of tendency, we look at the sSFR-mass
       relation per morphological type in the three density bins at
       $z\sim1$.  However, this analysis, due to the poor statistics
       of the current high z sample, can provide only an indication.
       As shown in Fig. \ref{early}, we observe a marginal increase of
       the mean sSFR of massive ($M/M_{\odot} > 5 \times 10^{11}$)
       early type galaxies from the low density to the high density
       bin. This figure is equivalent to the right panel of Fig. \ref{figura7} limited to the high mass bin. Late type and irregular galaxies do not exhibit any
       variation of the mean sSFR as a function of density given the
       current accuracy.  The fraction of massive early type galaxies
       varies from $\sim$ 20\% in the low density regime to $\sim$
       $40\%$ at the highest density at the expenses of the late type
       galaxy fraction which decreases from 40\% to 20\%, while the
       fraction of irregular galaxies remains constant. The tilt of
       the sSFR-mass relation at high density could be due to the
       presence of a larger number of massive and highly star forming
       early type galaxies. We point out that this is in agreement
       with the marginal increase of the density-sSFR relation
       observed in the high mass bin in the left panel of Fig.
       \ref{figura4b}. However, to confirm this indication we need a dataset with much higher statistics. In addition, the picture proposed here is partial because we can not check what is the role of the star forming galaxies with early-type morphology in the local SDSS star forming sample. Indeed the visual morphological classification applied to the high z star forming sample is just unfeasible for the large Sloan sample. The detailed study of the sSFR-mass relation as a function of the morphological type and environemnt in an homogeneous way and on a dataset with much higher statistics is the aim of another paper of this series.

\section{Discussion}

The main structures sampled by our high density regimes are two clusters in formation: a low mass cluster at $z=0.73$ at the center of two long filaments in the GOODS-S region, and two merging groups at $z=1.016$ in GOODS-N. Recently, Fadda et al. (2008) conducted a very wide field MIR observation with Spitzer MIPS which covers two filaments around the Abell 1763 cluster ($z= 0.23$) , and found that the fraction of starburst galaxies is more than twice larger in the filaments than in the inner region or outer fields of the clusters.  The enhancement of star-forming activity in filaments is also suggested in nearby clusters in optical studies (e.g. Porter \& Raychaudhury 2007; Porter et al. 2008).  Tzanavaris et al. (2010) find also an enhanced sSFR in members of gas rich Hickson compact groups (HGC) harboring mainly spirals and irregulars with respect to gas poor groups dominated by ellipticals and S0 galaxies.  Following Verdes-Montenegro et al. (2001) they argue that S/I and E/S0 HGC galaxies may constitute two distinct subclasses, consistent with being the two extremes of a possible evolutionary sequence progressing from the S/I high-sSFR subclass to the E/S0 low-SSFR one.  Initially, loose groups contract to a more compact configuration (Barton et al. 1998).  At this stage, most of the H I is found in galaxy disks, which constitute the prevailing morphological type. As the effects of tidal interactions gain in importance with time, an increasing fraction of the group H I mass is stripped from the interstellar medium of member galaxies and forms tidal tails, bridges, and intergalactic structures.  At this stage old, gas poor groups host quiescent elliptical and S0 galaxies.  This scenario is also in agreement with the more recent findings of Hwang et al. (2010) based on the cross-correlation of the SDSS galaxy catalog and IRAS faint source catalog. They find that the star formation activity in LIRGs and ULIRGs is triggered by galaxy galaxy interaction. Moreover, the star formation activity is boosted when the interaction happens between gas rich galaxies, while it is suppressed when a gas-rich galaxy encounter an early type gas poor system. 

Tran et al (2009) analyses the star formation activity of a super
    galaxy group at z $\sim$ 0.37. The system is merging to form a
    galaxy cluster comparable in mass to Coma. The super-group has a
    significantly higher fraction of dusty star-forming members than
    the massive cluster at the same redshift. The star forming galaxy
    population of the super-group is composed of luminous ($M_B
    <-20.5$) and relatively low mass ($10^{10} < M/M_{\odot} < 4
    \times 10^{10}$) members with late type morphology. Most of these
    members are outside the cluster core. Similarly, Vulcani et al.
    (2010) find that the median SFR of groups at $0.4<z<0.8$ is
    comparable to the median SFR of field galaxies, while in cluster
    SF galaxies they observe a much lower star formation activity.

A similar scenario is confirmed at redshift $z\sim 0.8$ by Koyama et al. (2008). They find that star-forming activity is enhanced in the intermediate-density cluster infalling region  between low-density general field and the high-density cluster core. They argue that the infalling galaxies would experience close interactions among themselves, even before they experience any significant influence of the gravitational potential of the cluster, and that of its ICM, on their rate of star formation. Particularly for galaxies falling along crowded filaments into a cluster, the local density of galaxies would already have begun to increase rapidly, and it could be appreciable at distances as large as twice the virial radius. Therefore, the most likely cause for enhanced star formation in the cluster infall region and groups is most likely due to galaxy$-$galaxy harassment, which is a rapidly acting process, working efficiently in crowded environments. Moving towards the center of the cluster, the infalling galaxies experience ram pressure stripping by the hot ICM and galaxy strangulation that quench the star formation rate.  At higher redshift, $z\sim 1.4-1.6$, Tran et al. (2010) and Hilton et al. (2010) observe enhanced star formation rates towards the cluster core and not only in the infalling region. This would suggest that at that epoch the processes quenching the star formation rate in the cluster highest density region did not yet take place or are not yet fully efficient.  

Putting our results in the context,  all tends to be consistent with the scenario described above. Indeed, we do observe segregation  of highly star forming galaxies, though AGN hosts, in the group density regime. In addition, the dynamical analysis of the structures considered here show that they are all groups with substructures, long filamentary structures and probably experiencing merging (Popesso et al. in prep.).  Thus, we are probably observing systems similar to the supergroup of Tran et al. (2009) at $z\sim 0.37$ with a confirmed enhanced star formation activity in the infalling regions.  However, our estimate of the galaxy local density is not accurate enough to distinguish between the infalling and virial regions. Thus, we can not confirm whether the reversal of the density-SFR relation still holds in the group core or whether it is dominated by the infalling systems.  In addition, our data provides an indication for a reversal of tendency at the high mass end of the sSFR-mass relation with respect to $z \sim 0.1$, possibly due to early type galaxies. This could indicate that we are whitnessing the formation of massive systems ($M/M_{\odot} > 10^{11}$) with early type morphology in the dense regions where they are usually segregated, as also observed at $z \sim 1.4-1.6$ (Tran et al. 2010, Hilton et al. 2010) . To have a clearer picture,  a more detailed study of the  members of the structures contained in our sample is needed and it is the aim of a second paper.

\section{Summary and Conclusions}

In order to build  a clean catalog of star forming galaxies at redshift $\sim$ 1, we complement the PACS catalogs in the GOODS fields  with MIPS 24 $\mu$m sources to cope with the different depth and incompletness levels in the GOODS-N and GOODS-S PACS observations. Applying additional cuts at 4.5 $\mu$m and at 24 $\mu$m we build a catalog complete in stellar mass down to $8 \times 10^9 M_{\odot}$ and $L_{IR}=10^{11}L_{\odot}$ with an average spectroscopic completeness of $80\%$ and above $60\%$ in any flux bin.  We identify X-ray detected AGNs and exclude broad line AGNs for which we can not obtain either a reliable estimate of the host stellar mass or star formation rate. We keep in the sample low luminosity obscured AGNs for which we obtain reliable estimates of the host galaxy properties. In order to build a reference sample of star forming galaxies at $z\sim 0$, we use the SDSS DR7 spectroscopic sample toghether with the SFR, stellar mass and classification of the MPA-JHU spectroscopic catalog.  We estimate the density parameter at high and low redshift in a homogeneous way. The density parameter is based on spectroscopic redshift information to properly identify low, intermediate and high density regimes.

We list here our findings:

\begin{itemize}

\item we analyse the density-SFR relation at redshift $\sim 1$ in two mass bins. We observe a reversal of the density-SFR relation, as previously found by Elbaz et al. (2007) and Cooper at al. (2008) for galaxies at masses $M/M_{\odot} > 5\times 10^{10}$. At lower masses the SFR and the density do not correlate. 
\item Thanks to the high accuracy provided by PACS in measuring the star formation rate also for AGN hosts (we have PACS detection for the 50\% most luminous AGN in the sample), we indentify in this class of objects the cause for the reversal of the global  density-SFR relation at $M/M_{\odot} > 8\times 10^9$.  This was not previously seen by  works based only Spitzer 24 $\mu$m data, which can be biased towards higher bolometric infrared luminosities by the AGN contribution (Elbaz et al. 2007) and works based on optical data which suffer even more by this bias (Cooper et al. 2008). 
\item Our data reveals that the fraction of AGN in the sample is rising towards high masses, high densities and high SFR. Thus, removing AGN from the sample is equivalent to removing an increasing fraction of the most star forming systems towards high densities, which causes the steepening of the density-SFR relation to disappear. 
\item Once AGN hosts are removed from the sample, as done for the local SDSS star forming sample based on optical data, the qualitative behavior of the density-SFR relation is consistent between high and low redshift.
\item We observe a clear density-sSFR anti-correlation which is due to the different contribution of massive and less massive galaxies, thus, to mass segregation. Removing the AGN hosts from the sample does not affect significantly the anti-correlation, since they are equally distributed in any bin of sSFR, thus, they have the same mean sSFR of star forming galaxies at the same mass.
\item The same trends and AGN effect are observed even when galaxies are separated in different morphological types.
\item We observe an indication for a reversal of tendency at the high mass end of the sSFR-mass relation with respect to the $z \sim 0.1$ relation. At $M/M_{\odot} > 10^{11}$ the mean specific star formation rate tends to be higher at higher density, while the opposite trend is observed in the local SDSS star forming sample.

\end{itemize}

\begin{acknowledgements}
PACS has been developed by a consortium of institutes led by MPE 
(Germany) and including UVIE (Austria); KUL, CSL, IMEC (Belgium); CEA, 
OAMP (France); MPIA (Germany); IFSI, OAP/AOT, OAA/CAISMI, LENS, SISSA 
(Italy); IAC (Spain). This development has been supported by the funding 
agencies BMVIT (Austria), ESA-PRODEX (Belgium), CEA/CNES (France),
DLR (Germany), ASI (Italy), and CICYT/MCYT (Spain).

We would like to thank the anonymous referee for the useful comments which significantly helped in improving the paper.

\end{acknowledgements}

\end{document}